\documentstyle[twocolumn,prl,aps,psfig]{revtex}
\begin{document}

\def\R{N}
\def\beqa{\begin{eqnarray}}
\def\eeqa{\end{eqnarray}}
\def\beqn{\begin{equation}}
\def\eeqn{\end{equation}}

\title{Concurrent computing machines and physical space-time}
\author{Philippe Matherat $^{(a)}$ and Marc-Thierry Jaekel $^{(b)}$}
\address{$^{(a)}$ Laboratoire Traitement et Communication de l'Information
\thanks{Laboratoire du CNRS et de l'Ecole Nationale Sup\'erieure des
T\'el\'ecommunications},
ENST, 46 rue Barrault, F75013 Paris France\\
$^{(b)}$ Laboratoire de Physique Th\'eorique
\thanks{Laboratoire du CNRS, de l'Ecole Normale Sup\'erieure et de
l'Universit\'e Paris Sud},
ENS, 24 rue Lhomond, F75005 Paris France
}
\date{LPTENS 01/05}

\maketitle

\begin{abstract}
Concrete computing machines, either sequential or concurrent, 
rely on an intimate relation between computation and time.
We recall the general characteristic properties 
of physical time and of present realizations of computing systems.
We emphasize
the role of computing interferences, i.e. the necessity 
to avoid them in order to give a causal implementation to logical operations.
We compare synchronous and
asynchronous systems, and make a brief survey of some methods used 
to deal with computing interferences.
Using a graphic representation, 
we show that synchronous and asynchronous circuits reflect the 
same opposition as the Newtonian and relativistic causal structures
for physical space-time.
\end{abstract}


\section{Introduction}

Are concurrent computing machines equivalent to Turing machines?
This question, which amounts to confront two fundamental notions
like time and computation may be treated in a purely
mathematical framework. Practical consequences however cannot be
independent of concrete realizations, that is
concrete machines performing actual computations in
physical time.

This remark may seem curious, if one aims at showing theorems, which cannot
depend on the physical properties of time or machines.
But, even in a mathematical treatment of concurrent computation,
one needs a representation of time. Usually, time is modelized as a real
parameter, shared by all parts of the computation.
Unfortunately, such a representation does not correspond to the observable
time that can be obtained from physical systems like clocks,
 neither to the reference time
that is defined by metrology, nor to the operational time that occurs in
practical realizations of logical circuits.
Without questioning the validity of
demonstrated theorems, difficulties may emerge
when trying to find
practical applications.

The distinction just made between abstract and concrete machines raises related
questions. When a machine $M$ can be simulated by a program $P$
running on another machine, how can one identify the concrete
machine $M$ and the program $P$?
 And in case a machine cannot be simulated on another one,
indicating some greater expressive power, is the latter
due to computation or to some fundamental physical property?
Before entering such questions, one must first consider
the sequential machines that are presently realized.
These machines which we get from a constructor, which are made of matter,
which transform electric energy into heat, which we communicate with
through a keyboard and a screen, why do we need them?
Let us note that all the features that make them concrete are inconveniences:
we would prefer them lighter, smaller, less power consuming and less
dissipating. It would be ideal to make all these parameters equal to zero.
In fact, we need them for their logical function, their ability to
compute. But then, since this function is mathematically known,
modelized and even simulated, where is the need for a concrete machine,
whose features are mainly inconveniences? The natural answer is
that these machines compute faster than humans can do, with
just a pencil and paper. And yet, pencil and paper are already
rudimentary elements of a concrete machine, using physical
objects to memorize different steps of computations.
The interest in concrete machines comes from their intimate relation
with physical time.

If computing machines go faster than humans, then one must be confident
in their action, as in most cases one is unable to check their output.
Indeed, in very specific cases, one can verify the correctness of a result
in much less time than is needed to obtain it.
This is the case for instance of the prime factorization of integers.
But few concrete applications have this property. In most cases,
one cannot check the result in much less time than the computation itself.
If the result is important, and no other way is available to obtain it,
then one must be confident in the machine.

What can support such a confidence? Necessarily reasoning, founded
on correct functioning of the machine at a given time on some
particular computations, generalized to other times and
other computations. A computing machine cannot be tested for all
computations it can do, at any time.
Even for a finite machine, the number of possible computations increases
exponentially with the memory size, and a memory of one hundred bits
already allows a number of configurations that cannot be tested in
less time than the age of the universe.
To establish a reasoning leading to confidence, one must:

- check that each elementary component effectively realizes the function
it has been designed for (physical validation).

- prove in a deductive way that the particular composition
of these elementary components building the machine effectively leads to
the global function used (logical validation).

The first condition is ensured by choices in implementation design
and by tests made by the constructor.
The second condition is obtained from a mathematical representation
of the machine
and from the logics of computation.
These two steps of validation require good representations of all components
at the physical level, and of the global machine at the logical level.
If the confidence one can put in a machine relies on good modelizations
of both its physical and logical functioning,
how could such a machine perform more than it has been designed for, more than
our present theories can modelize?
Even if the existence of a new type of calculus, still unknown today,
can be envisaged, with machines performing this new type of calculus, 
how could 
one build such machines without having for them good modelizations?
In such a case, one could not ensure the two validation steps,
and one could not say that these machines operate correctly
 neither be confident
in their ouput.

One consequence is that realistic models are necessary, both of
the computational structure and of the physical implementation of
logical operators.
Modelization is made easier when logical and physical constraints can be
separated. This is the reason for developing 
sequential machines or synchronous concurrent
machines. In that case, the logical
validation of the machine can be made, whilst
ignoring the implementation characteristics
of its components. The latter will finally and
mainly limit the performance of the machine
through the value of the clock frequency.
The machine can equivalently be simulated on another concrete machine
with identical clock frequency, at the expense of slower performances.
However, in the case of asynchronous concurrent machines, 
logical and physical constraints are more involved.
Although machines built with asynchronous circuits are less widely used,
much effort has been devoted to their understanding and modelization
\cite{Seitz,Sutherland-Micropipelines,Martin-Hoare,Ebergen-Dist.Comp.,Davis-Nowick,Mallon-Udding-Verhoeff}.
In fact, they may even appear as an unavoidable evolution of computing machines.
On one hand, clock timed circuits are reaching limits where
clock signal distribution consumes too many resources and
progress in performances approaches saturation point.
On the other hand, asynchronous circuits
constitute the most general class of circuits, and thus
allow one to express in the most general way the questions raised by
the implementation of computation on physical systems,
and the solutions that may be brought.

In this article, we shall be concerned with 
the relation of concrete computing machines with physical time.
After recalling the general characteristic properties of physical time and 
of computing machines which are presently realized, we shall compare
the solutions provided by synchronous and asynchronous systems to the 
implementation of logical operations. We shall show that they give different
implementations of causal relations, reflecting in that way different causal 
structures for space-time.

\section{Physical time}

The notion of time may be seen to follow from two necessities. From a logical
point of view, time can be considered as the concept which allows one to make
a distinction between two
different types of propositions: general and universal propositions
(like mathematical ones) which are eternal, and particular propositions
which are related to changing reality (like those describing
physical systems). Moreover, time is also rendered necessary by the formulation
of physics: time is the concept which allows one to give a formal expression to
movement, and hence to the laws of physics.

Properties of time are in fact imposed by the functions
that this notion must fulfill.
From the logical side, time allows one to conceive a same object
by characterizing it by its different states,
 these states being asssociated with
the object at different times.
A time parameter can then be used not only to index the
different states characterizing a same object, but also to organize
the states of different objects into classes of simultaneity.

The relation of order that can be introduced on the time parameter
allows one to define a relation of logical causality between
the state transitions affecting different objects. However, in order to
be realized physically, for instance on real machines, the causal relation
between states cannot be independent of the real motions affecting physical
systems. In particular, the simultaneity classes defined with the help of
the time parameter must coincide  with those that are associated with
real events occuring in physical space, hence with the physical time.

The notion of physical time is intimately related to the laws of physics.
After having remarked that pendulum oscillations are isochronous,
Galileo Galilei could give a
mathematical representation of motion induced by free fall, by relating the
undergone distance to the elapsed time,
the latter being understood as a universal
reference for all motions.
The existence of such a reference is made possible by the existence of
physical laws governing all movements, and in particular by the existence
of regular movements like inertial motions.

This introduction of time leaves an important conventional part
in the definition of a time reference, even if a natural choice is provided by
motions which appear as most regular, like the Earth motion around the Sun.
This leads in fact to distinguish two types of time. Thus,
Leibniz \cite{Russell-Phi.Leib.},
relying on logical arguments, could consider that
space and time are mere relations between objects or events, which are
fixed by an observer in a conventional way, thus building subjective space
and time. Still, one is also bound to admit the existence of
objective space and time, as the only way to understand how physical laws
governing displacements of objects and time ordering of events can
be formulated in a universal way, independently of the observer.

The formulation of the universal law of gravitation led Newton \cite{Newton}
to fix the role played by time in physical laws, and to endow it with
the mathematical
representation that we still use nowadays: that of a real parameter which
all physical quantities depend on. In fact, Newton introduced two different
notions of time, which he distinguished both in their conception and in their
usage. The first one, which he called "absolute and mathematical",
allowed him to write mathematical equations for the laws of mechanics
and gravitation. The second notion, which he called "common and sensible",
allowed him to relate the motions of different physical systems, including
clocks. Even if Newton privileged the first notion, which he considered
as representing absolute space and time, seeing clocks as systems
to be improved in order to make them as close as 
possible to ideal space and time,
he nevertheless made two distinct uses of these notions. The first one, which
identifies with the curvilinear coordinate on the planet's trajectory,
he used
as a mathematical tool to deal with infinitesimals of different orders.
The second one, which is the physical time as can be
 defined by Kepler's area law,
he used as a measure of inertial motions,
which he compared planetary motions with.

The theory of relativity \cite{Einstein} has led to question
the a priori and absolute character of physical space and time.
According to relativity, the notion of time 
relies on clocks, the date of an event being defined by coincidence
of this event with a top delivered by a clock located at the same place.
But in order to be defined in whole space, the notion of time also relies on
the exchange of light signals, which
are necessary to compare and synchronize the indications of
remote clocks.
The universal and finite velocity of light propagation then leads
to a definition of time simultaneity which depends on the observer's motion.
In other words, time simultaneity is not given a priori
but results from a construction, or clock synchronization.
By exchanging light signals, on which time references
provided by clocks are encoded, one can compare these references
and synchronize clocks.
Then, time allows one to construct space.
By comparing the light signals received from several remote clocks, one can,
by quadrangulation, determine positions both in time and space.
This relativistic definition of time and space is rendered necessary as soon
as a high precision must be attained. This is the case for instance when
corrections linked to the finite velocity of light, or relativistic effects,
must be taken into account \cite{Landau-Lifschitz,Wrinkler}.
Hence, this relativistic definition is the one
used in physics for high precision space-time measurements
\cite{Vessot}, and in metrology to define time and space standards
\cite{Quinn} and to construct the
space-time reference systems required by physics \cite{Petit,Wolf}.
It is also the one
used in modern practical positioning systems at the surface of the Earth,
like GPS \cite{Lewandowski-Thomas,Leschiutta}.
Finally, as it will appear in the following, it is also the
notion of time which is implicitly used by asynchronous communicating
and computing systems \cite{Lamport-Clocks}.

The consequences of the theory of relativity on our conception of
space and time have been remarkably
discussed at the logical level by Russell \cite{Russell-ABC-Rel.,Russell-AM}.
Our representation in terms of permanent material structures located in space
and evolving according to a unique external time, must be replaced by that
in terms of events which are located both in space and time.
This conception of space-time not only affects the
formulation of modern
theories in a fundamental way \cite{Jaekel-Reynaud}, but also
underlies present applications in physics and metrology \cite{Guinot}.

When refering to physical time, simultaneity classes cannot be defined a priori
any more, and rely on a physical implementation by means of propagating light
signals. This constructive character of time has important consequences on the
functioning of devices which rely on the physical exchange of information.
Causal relations between events cannot be derived by simple comparison
with an external, a priori given, parameter.
For systems which are unlocalized in space, like communicating processors,
this means that the time order relation of occuring events, even if it
can be defined unambiguously  at the local level of each processor, nevertheless
requires a more complete representation to
be defined over the whole system in a consistent way
\cite{Seitz,Sutherland-Micropipelines}.
In the following, we shall analyse how the functioning of actual devices
depends on the causal structure of physical space-time.

\section{Logical devices}

To discuss the intimate relation between time and computation, 
one must first recall some general principles which
underly the physical implementation of computing systems,
and which are applied in concrete machines 
realized with present technologies.

\bigskip
\noindent {\bf 3.1 Implementation of logical operations}

\noindent In CMOS (Complementary MOS) technology, 
logical gates are implemented using two 
electrical networks $\R_u$ and $\R_d$, as represented on Figure~1.
$x_i$ describe input channels, and $z$ the output channel.
$\R_u$ and $\R_d$ are built with electrical switches 
which are combined in series/parallel networks,
thus allowing to implement the logics of propositions \cite{Shannon}.

\bigskip
\begin{picture}(0,0)%
\psfig{file=push-pull.pstex}%
\end{picture}%
\setlength{\unitlength}{4144sp}%
\begingroup\makeatletter\ifx\SetFigFont\undefined%
\gdef\SetFigFont#1#2#3#4#5{%
  \reset@font\fontsize{#1}{#2pt}%
  \fontfamily{#3}\fontseries{#4}\fontshape{#5}%
  \selectfont}%
\fi\endgroup%
\begin{picture}(1530,2808)(856,-2635)
\put(2341,-1051){\makebox(0,0)[lb]{\smash{\SetFigFont{12}{14.4}{\rmdefault}{\mddefault}{\updefault}Output}}}
\put(1711,-826){\makebox(0,0)[lb]{\smash{\SetFigFont{12}{14.4}{\rmdefault}{\mddefault}{\updefault}$N_{u}$}}}
\put(1036,-1816){\makebox(0,0)[lb]{\smash{\SetFigFont{12}{14.4}{\rmdefault}{\mddefault}{\updefault}$x_{i}$}}}
\put(856,-1276){\makebox(0,0)[lb]{\smash{\SetFigFont{12}{14.4}{\rmdefault}{\mddefault}{\updefault}Inputs}}}
\put(2386,-1276){\makebox(0,0)[lb]{\smash{\SetFigFont{12}{14.4}{\rmdefault}{\mddefault}{\updefault}$z$}}}
\put(1711,-1816){\makebox(0,0)[lb]{\smash{\SetFigFont{12}{14.4}{\rmdefault}{\mddefault}{\updefault}$N_{d}$}}}
\put(1396,-151){\makebox(0,0)[lb]{\smash{\SetFigFont{12}{14.4}{\rmdefault}{\mddefault}{\updefault}(logical true)}}}
\put(1036, 29){\makebox(0,0)[lb]{\smash{\SetFigFont{12}{14.4}{\rmdefault}{\mddefault}{\updefault}Positive voltage source}}}
\put(1036,-2401){\makebox(0,0)[lb]{\smash{\SetFigFont{12}{14.4}{\rmdefault}{\mddefault}{\updefault}Negative voltage source}}}
\put(1396,-2581){\makebox(0,0)[lb]{\smash{\SetFigFont{12}{14.4}{\rmdefault}{\mddefault}{\updefault}(logical false)}}}
\put(1036,-781){\makebox(0,0)[lb]{\smash{\SetFigFont{12}{14.4}{\rmdefault}{\mddefault}{\updefault}$x_{i}$}}}
\end{picture}

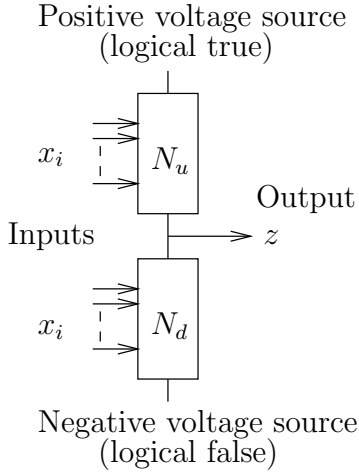
Fig. 1. A logical gate.

\bigskip
Each switch is implemented with a transistor. A function implemented
by $\R$ will be said to be true, for a particular set of values of $x_i$,
if and only if there is a path that connects
extreme connections
of $\R$.
For instance, for the network represented by Figure~1:              

- when $\R_u$ is {\it true},
output $z$ is forced to value {\it true}.

- when $\R_d$ is {\it true},
output $z$ is forced to value {\it false}.

The possibility that ($\R_u = \R_d = true$)
for some values of $x_i$ must be
excluded, otherwise current could flow both through $\R_u$ and $\R_d$,
resulting in a short circuit between voltage sources. In the following, we
shall always impose that
($\lnot \R_u \lor \lnot \R_d$)
is verified for all configurations of variables
$x_i$ ($\lnot$, $\lor$ and $\land$ define as usual negation, logical 
disjunction and logical conjunction).

\bigskip
\begin{picture}(0,0)%
\psfig{file=nand.pstex}%
\end{picture}%
\setlength{\unitlength}{4144sp}%
\begingroup\makeatletter\ifx\SetFigFont\undefined%
\gdef\SetFigFont#1#2#3#4#5{%
  \reset@font\fontsize{#1}{#2pt}%
  \fontfamily{#3}\fontseries{#4}\fontshape{#5}%
  \selectfont}%
\fi\endgroup%
\begin{picture}(3117,2025)(1126,-2401)
\put(2476,-1141){\makebox(0,0)[lb]{\smash{\SetFigFont{12}{14.4}{\rmdefault}{\mddefault}{\updefault}$z$}}}
\put(1801,-2401){\makebox(0,0)[lb]{\smash{\SetFigFont{12}{14.4}{\rmdefault}{\mddefault}{\updefault}0}}}
\put(1306,-1861){\makebox(0,0)[lb]{\smash{\SetFigFont{12}{14.4}{\rmdefault}{\mddefault}{\updefault}$x_{1}$}}}
\put(4006,-781){\makebox(0,0)[lb]{\smash{\SetFigFont{12}{14.4}{\rmdefault}{\mddefault}{\updefault}$z$}}}
\put(1306,-1501){\makebox(0,0)[lb]{\smash{\SetFigFont{12}{14.4}{\rmdefault}{\mddefault}{\updefault}$x_{2}$}}}
\put(1126,-826){\makebox(0,0)[lb]{\smash{\SetFigFont{12}{14.4}{\rmdefault}{\mddefault}{\updefault}$x_{1}$}}}
\put(1666,-826){\makebox(0,0)[lb]{\smash{\SetFigFont{12}{14.4}{\rmdefault}{\mddefault}{\updefault}$x_{2}$}}}
\put(3151,-646){\makebox(0,0)[lb]{\smash{\SetFigFont{12}{14.4}{\rmdefault}{\mddefault}{\updefault}$x_{1}$}}}
\put(1531,-511){\makebox(0,0)[lb]{\smash{\SetFigFont{12}{14.4}{\rmdefault}{\mddefault}{\updefault}1}}}
\put(2026,-511){\makebox(0,0)[lb]{\smash{\SetFigFont{12}{14.4}{\rmdefault}{\mddefault}{\updefault}1}}}
\put(3151,-1051){\makebox(0,0)[lb]{\smash{\SetFigFont{12}{14.4}{\rmdefault}{\mddefault}{\updefault}$x_{2}$}}}
\end{picture}

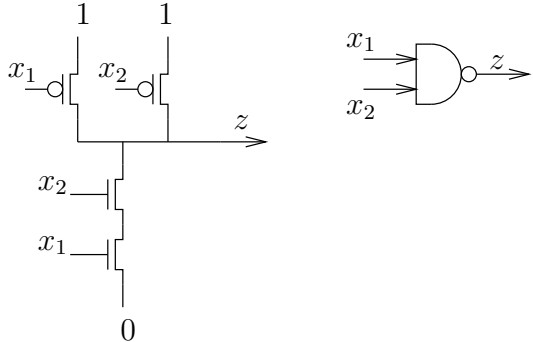
Fig. 2. Nand gate.

\bigskip
Two cases must be considered.

1- $\R_u$ and $\R_d$ are always opposite, for all values of variables $x_i$,
($\R_u = \lnot \R_d$).
This case implements propositions of classical logic
(Complementary MOS). The simple example of the {\it nand} gate
is represented on Figure~2 (a bubble represents negation):
\begin{eqnarray}
z = \R_{u} &=&\lnot x_{1} \lor \lnot x_{2} \nonumber\\
\R_{d} &=& \lnot \R_u = x_{1} \land x_{2}
\end{eqnarray}

2- $\R_u$ and $\R_d$ can be simultaneously {\it false}, i.e. 
($\lnot \R_u \land \lnot \R_d$) can be {\it true}.
In such configurations, the output $z$ is not connected to any voltage source.
Then, because of electrical capacities, $z$ memorizes its previous value.
This allows one to realize memories, like the {\it latch}
represented in Figure~3:
\begin{eqnarray}
&&\R_{u}= x \land e \nonumber\\
&&\R_{d}= \lnot x \land e
\end{eqnarray}

\bigskip
\begin{picture}(0,0)%
\psfig{file=latche.pstex}%
\end{picture}%
\setlength{\unitlength}{4144sp}%
\begingroup\makeatletter\ifx\SetFigFont\undefined%
\gdef\SetFigFont#1#2#3#4#5{%
  \reset@font\fontsize{#1}{#2pt}%
  \fontfamily{#3}\fontseries{#4}\fontshape{#5}%
  \selectfont}%
\fi\endgroup%
\begin{picture}(1620,1179)(631,-2761)
\put(2251,-1951){\makebox(0,0)[lb]{\smash{\SetFigFont{12}{14.4}{\rmdefault}{\mddefault}{\updefault}$z$}}}
\put(946,-1951){\makebox(0,0)[lb]{\smash{\SetFigFont{12}{14.4}{\rmdefault}{\mddefault}{\updefault}$x$}}}
\put(946,-2536){\makebox(0,0)[lb]{\smash{\SetFigFont{12}{14.4}{\rmdefault}{\mddefault}{\updefault}$e$}}}
\put(1981,-1726){\makebox(0,0)[lb]{\smash{\SetFigFont{12}{14.4}{\rmdefault}{\mddefault}{\updefault}Data output}}}
\put(721,-2761){\makebox(0,0)[lb]{\smash{\SetFigFont{12}{14.4}{\rmdefault}{\mddefault}{\updefault}Enable}}}
\put(1576,-2131){\makebox(0,0)[lb]{\smash{\SetFigFont{12}{14.4}{\rmdefault}{\mddefault}{\updefault}L}}}
\put(631,-1726){\makebox(0,0)[lb]{\smash{\SetFigFont{12}{14.4}{\rmdefault}{\mddefault}{\updefault}Data input}}}
\end{picture}
 
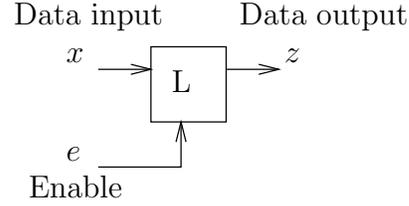
Fig. 3. Transparent latch.

\bigskip
In practice, in order to have memorization last for long enough,
 and quite generally for
all memories, one must ensure memory stability by using some feed-back, by means
of a looped amplifier. This feed-back can be permanent (static) or 
recurrent (dynamic logic). One possibility of electrical feed-back is shown
in Figure~4, where two looped amplifiers have been added on output $z$, one
of them being weak, in the sense that it cannot create any serious
short circuit when it conflicts with any of the two networks $\R_u$ and $\R_d$. 

\bigskip

\begin{picture}(0,0)%
\psfig{file=push-pull-mem.pstex}%
\end{picture}%
\setlength{\unitlength}{4144sp}%
\begingroup\makeatletter\ifx\SetFigFont\undefined%
\gdef\SetFigFont#1#2#3#4#5{%
  \reset@font\fontsize{#1}{#2pt}%
  \fontfamily{#3}\fontseries{#4}\fontshape{#5}%
  \selectfont}%
\fi\endgroup%
\begin{picture}(1992,2385)(1126,-2401)
\put(1801,-151){\makebox(0,0)[lb]{\smash{\SetFigFont{12}{14.4}{\rmdefault}{\mddefault}{\updefault}1}}}
\put(1126,-781){\makebox(0,0)[lb]{\smash{\SetFigFont{12}{14.4}{\rmdefault}{\mddefault}{\updefault}$x_{i}$}}}
\put(1711,-871){\makebox(0,0)[lb]{\smash{\SetFigFont{12}{14.4}{\rmdefault}{\mddefault}{\updefault}$N_{u}$}}}
\put(1711,-1816){\makebox(0,0)[lb]{\smash{\SetFigFont{12}{14.4}{\rmdefault}{\mddefault}{\updefault}$N_{d}$}}}
\put(1801,-2401){\makebox(0,0)[lb]{\smash{\SetFigFont{12}{14.4}{\rmdefault}{\mddefault}{\updefault}0}}}
\put(2566,-826){\makebox(0,0)[lb]{\smash{\SetFigFont{10}{12.0}{\rmdefault}{\mddefault}{\updefault}weak}}}
\put(1126,-1816){\makebox(0,0)[lb]{\smash{\SetFigFont{12}{14.4}{\rmdefault}{\mddefault}{\updefault}$x_{i}$}}}
\put(2971,-1141){\makebox(0,0)[lb]{\smash{\SetFigFont{12}{14.4}{\rmdefault}{\mddefault}{\updefault}$z$}}}
\end{picture}

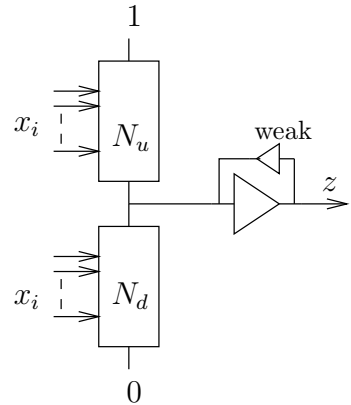
Fig. 4. Electrical feed-back.

\bigskip
Another way to realize a stable memory is to implement a static feed-back on
a logical gate corresponding to the first case. 
Then, the quoted {\it latch} can be realized with a looped multiplexer 
({\it mux}),
 as shown 
in Figure~5:
\begin{eqnarray}
\label{latchdef}
z=N_{u} &=&(x \land e) \lor (z \land \lnot e)\nonumber\\
N_{d} &=& \lnot N_{u}
\end{eqnarray}
\begin{picture}(0,0)%
\psfig{file=latche-mux.pstex}%
\end{picture}%
\setlength{\unitlength}{4144sp}%
\begingroup\makeatletter\ifx\SetFigFont\undefined%
\gdef\SetFigFont#1#2#3#4#5{%
  \reset@font\fontsize{#1}{#2pt}%
  \fontfamily{#3}\fontseries{#4}\fontshape{#5}%
  \selectfont}%
\fi\endgroup%
\begin{picture}(1632,1497)(991,-2491)
\put(2611,-1501){\makebox(0,0)[lb]{\smash{\SetFigFont{12}{14.4}{\rmdefault}{\mddefault}{\updefault}$z$}}}
\put(991,-1681){\makebox(0,0)[lb]{\smash{\SetFigFont{12}{14.4}{\rmdefault}{\mddefault}{\updefault}$x$}}}
\put(991,-2491){\makebox(0,0)[lb]{\smash{\SetFigFont{12}{14.4}{\rmdefault}{\mddefault}{\updefault}$e$}}}
\put(1711,-1186){\makebox(0,0)[lb]{\smash{\SetFigFont{12}{14.4}{\rmdefault}{\mddefault}{\updefault}Mux}}}
\end{picture}

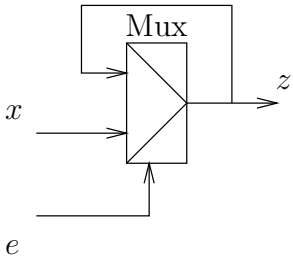
Fig. 5. Transparent latch with multiplexer.

\bigskip
This exhibits a very general difficulty which characterizes looped systems:
variable $z$ appears on both sides of its defining equation (\ref{latchdef}).
This equation does not mean that an equality must be
realized, for instance with electric voltages, but that the following
assignment must be realized:
\begin{equation}
\label{latcha}
z \leftarrow \left(x \land e \right) \lor \left(z \land \lnot e\right)
\end{equation}
\begin{equation}
\label{latche}
z_a = \left(x \land e \right) \lor \left(z_b \land \lnot e\right)
\end{equation}
In other words, two values of the variable $z$ must be distinguished, which 
correspond to successive times: $z_a$ (after) and $z_b$ (before) are linked
by equation (\ref{latche}).
It is required that the two values $z_a$ and $z_b$ do not interfere,
and that variable $z$ change from $z_b$ to $z_a$.
The assignment represented by equation (\ref{latcha}) expresses a causality
requirement that must be implemented in order to realize computations.

In the particular case of the {\it latch} just described, 
operation may only cause problem
in case $e$ is falling. Indeed, in other cases: 

- when $e$ is low, $z$ is memorized

- when $e$ is high, $z$ copies $x$

- when $e$ rises, $z$ begins to copy $x$

so that the circuit operates correctly in these three cases.
However, if $e$ falls down while $x$ changes, $z$ will hesitate between two
values of $x$. The whole circuit may enter a metastable state which is invalid
(electric voltage will stay in metastable balance at an intermediate level)
and which may last for an unbounded time. It may
leave this state for any of the two possible values of $z$, and this
in an undeterministic way, which may not be eventually acceptable for the type
of computation envisaged.
Let us note that the circuit of Figure~4 shows the same defects, for it 
involves a feed-back, although this may be less apparent when treated at 
the electrical level.

Although chosen here as an example, the {\it latch} shows properties which are 
encountered quite generally in looped devices.
This brief discussion shows that the correct operation of a circuit
cannot be analysed without taking its environment into account, 
in particular the time ordering relations of input and output signals.
This is entailed by the existence of loops 
and must be dealt with quite generally,
for computing machines are naturally looped systems.
In all cases, time constraints must be implemented  
in order to ensure the causality relations which are necessary for computation.
These constraints will take very different forms, according to the type
of implementation chosen, whether by means of
synchronous or asynchronous systems.
Before discussing separately these two classes of systems, we shall first recall
one important property they share, as it is also imposed by implementation
of complex computations, the property of modularity.

\bigskip
\noindent {\bf 3.2 Modularity}

\noindent There exist many various ways to organize electronic components into
logical circuits, in order to realize machines performing computations.
Usually and quite generally, one defines
complex circuits as hierarchies built with 
elementary circuits called primitives. This method, imposed by practical
considerations, indeed hints at a logical necessity: one must be able to
design and realize with the same
rigour circuits of increasing complexity. More precisely, one must
insure that circuits implementing logical functions of high complexity level
behave as they should, and one must obtain this confidence in a rather
short time. Because of the exponential increase of the number of
configurations to check, this requirement implies that
a direct physical test of
the circuit's behaviour soon becomes impossible
when the complexity of the logical function increases.
This aim can then only be attained with the help of modular implementations,
by taking advantage both of their composite logical structure, and of the
logical simplicity of chosen primitives \cite{Matherat-Jaekel}.
Proofs relying on known properties
of composition of primitives may be developed, which allow one to deduce
the correct functioning
of a whole modular complex from that of its constituent primitives.
Then, a test of the whole complex reduces to that of some of its
constituents, which are logically simple.
Although efficient, such strategy may not reveal itself so straightforward.
According to the type of physical implementation chosen for the primitives,
problems may appear which prevent the systematic development of
complex circuits operating correctly, and which
do not occur when the choice of primitives is modified,
or when pecular constraints are put on their composition.
Then, there results that logical and physical aspects of modular
implementations must be analysed concurrently.

\section{Synchronous and asynchronous circuits}

Time appears in computing systems very early,
already in the definition of the electronic circuits
which implement logical functions. Most circuits which are known and used are
synchronous. Synchronous circuits may be defined as automata whose
transitions between successive states are triggered by pulses delivered by
a global clock. 
An alternative class of circuits is provided by asynchronous 
circuits.
In this section, we introduce the strategies followed by these two 
main classes of circuits for making the implementation of computation 
effective, and, in particular, for dealing with problems of
computation interferences.

\bigskip
\noindent {\bf 4.1 Synchronous circuits}

\noindent VLSI circuits which are produced nowadays 
are highly concurrent devices
(a microprocessor can contain up to
$10^7$ transistors, i.e. $10^6$ logical gates),
and yet most of them can be modelized as
non concurrent devices and can be considered
as single finite automata. This property comes from their synchronous
 character, which means that all operations on internal memories are
simultaneously activated by a single pulse of a global clock shared
by the whole circuit. 

Synchronous implementations use a global clock to avoid the stability problem
which has been discussed in previous section. More precisely, with
same notations, all {\it latches} are systematically
operated in such a way to ensure that logical variables $x$ be stable when
variables $e$ are falling. There exist many different types of memories,
but all present the same problem, reflecting the time character of
logical assignment. For the sake of simplicity, we shall only  discuss the case
of the previous {\it latch}, and consider it as a generic example.
A synchronous device using two items of this {\it latch} for each register bit
(master-slave {\it flip-flop})
is sketched on Figure~ 6.
The enabling signals $e_1$ and $e_2$ are mutually excluded in time, and are
derived systematically from a common clock. The output is fed back
under the form of input variables $Q_i$
into a combinatorial operator.
If the clock period is larger than the feedback time, then
variables $Q^\prime_i$ are always stable
when $e_1$ is falling and the {\it latches} act as
required, i.e., they make the iteration of the combinatorial function
effective.

The global state $Q$ is encoded by
the state of all memory bits, and can only change
at the arrival of a clock pulse.
Regarding specification and design, synchronous circuits may be considered as
modular composites, where primitives, and other modules as well, are finite
automata of the type described by Figure~6.
The register encodes the state of the circuit, while the combinatorial
operator represents the implementation of its transition function.
All registers are activated by a single clock pulse. By connecting
several automata of this type, one obtains another automaton of the same
type, only with a larger memory and a more complex combinatorial function.
In such a representation, and from a logical point of view,
neither time nor space are involved. One only needs to consider the
successive logical steps associated with successive clock periods.
The logical time of a computation reduces to a mere integer, 
which one only relates to physical time by multiplying it
by the mean clock period.
In other words, time is discretized.

\bigskip

\begin{picture}(0,0)%
\psfig{file=synchrone.pstex}%
\end{picture}%
\setlength{\unitlength}{4144sp}%
\begingroup\makeatletter\ifx\SetFigFont\undefined%
\gdef\SetFigFont#1#2#3#4#5{%
  \reset@font\fontsize{#1}{#2pt}%
  \fontfamily{#3}\fontseries{#4}\fontshape{#5}%
  \selectfont}%
\fi\endgroup%
\begin{picture}(3072,5916)(946,-6010)
\put(1711,-3031){\makebox(0,0)[lb]{\smash{\SetFigFont{12}{14.4}{\rmdefault}{\mddefault}{\updefault}$i$-th bit of register}}}
\put(2926,-2401){\makebox(0,0)[lb]{\smash{\SetFigFont{12}{14.4}{\rmdefault}{\mddefault}{\updefault}Clock}}}
\put(3331,-1411){\makebox(0,0)[lb]{\smash{\SetFigFont{12}{14.4}{\rmdefault}{\mddefault}{\updefault}$Q$}}}
\put(946,-1951){\makebox(0,0)[lb]{\smash{\SetFigFont{12}{14.4}{\rmdefault}{\mddefault}{\updefault}Inputs}}}
\put(2746,-1411){\makebox(0,0)[lb]{\smash{\SetFigFont{12}{14.4}{\rmdefault}{\mddefault}{\updefault}$Q\prime$}}}
\put(3241,-2176){\makebox(0,0)[lb]{\smash{\SetFigFont{12}{14.4}{\rmdefault}{\mddefault}{\updefault}$e_{2}$}}}
\put(2746,-646){\makebox(0,0)[lb]{\smash{\SetFigFont{12}{14.4}{\rmdefault}{\mddefault}{\updefault}Register}}}
\put(2881,-2176){\makebox(0,0)[lb]{\smash{\SetFigFont{12}{14.4}{\rmdefault}{\mddefault}{\updefault}$e_{1}$}}}
\put(1666,-1276){\makebox(0,0)[lb]{\smash{\SetFigFont{10}{12.0}{\rmdefault}{\mddefault}{\updefault}Combinatorial}}}
\put(1846,-1411){\makebox(0,0)[lb]{\smash{\SetFigFont{10}{12.0}{\rmdefault}{\mddefault}{\updefault}function}}}
\put(3511,-1816){\makebox(0,0)[lb]{\smash{\SetFigFont{12}{14.4}{\rmdefault}{\mddefault}{\updefault}Outputs}}}
\put(1621,-5281){\makebox(0,0)[lb]{\smash{\SetFigFont{12}{14.4}{\rmdefault}{\mddefault}{\updefault}$e_{2}$}}}
\put(1621,-4921){\makebox(0,0)[lb]{\smash{\SetFigFont{12}{14.4}{\rmdefault}{\mddefault}{\updefault}$e_{1}$}}}
\put(3376,-5551){\makebox(0,0)[lb]{\smash{\SetFigFont{12}{14.4}{\rmdefault}{\mddefault}{\updefault}time}}}
\put(1936,-5956){\makebox(0,0)[lb]{\smash{\SetFigFont{12}{14.4}{\rmdefault}{\mddefault}{\updefault}One clock period}}}
\put(3376,-3391){\makebox(0,0)[lb]{\smash{\SetFigFont{12}{14.4}{\rmdefault}{\mddefault}{\updefault}$Q_{i}$}}}
\put(1261,-3391){\makebox(0,0)[lb]{\smash{\SetFigFont{12}{14.4}{\rmdefault}{\mddefault}{\updefault}$Q\prime _{i}$}}}
\put(1936,-3616){\makebox(0,0)[lb]{\smash{\SetFigFont{12}{14.4}{\rmdefault}{\mddefault}{\updefault}L}}}
\put(2746,-3616){\makebox(0,0)[lb]{\smash{\SetFigFont{12}{14.4}{\rmdefault}{\mddefault}{\updefault}L}}}
\put(2116,-4201){\makebox(0,0)[lb]{\smash{\SetFigFont{12}{14.4}{\rmdefault}{\mddefault}{\updefault}$e_{1}$}}}
\put(2926,-4201){\makebox(0,0)[lb]{\smash{\SetFigFont{12}{14.4}{\rmdefault}{\mddefault}{\updefault}$e_{2}$}}}
\end{picture}

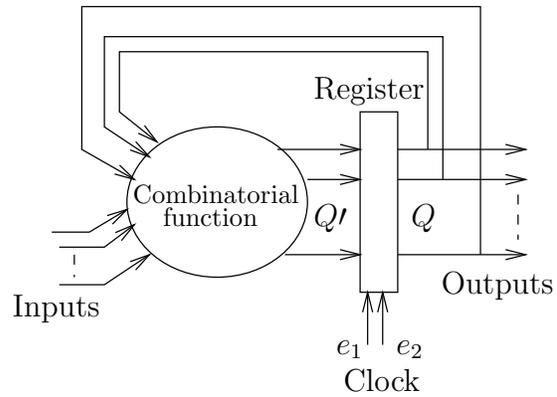
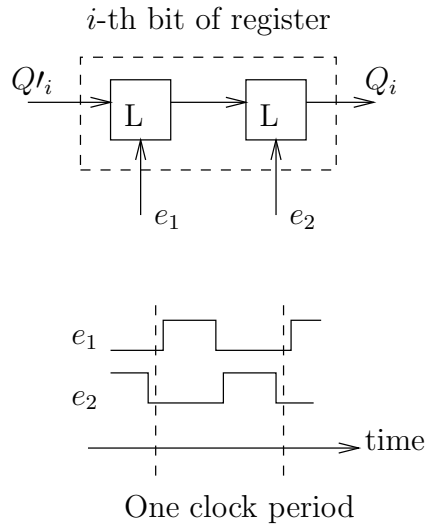
Fig. 6. Synchronous circuit.

\bigskip
The only physical constraint one must impose is that the clock frequency be
smaller than the limit value necessary for all internal propagations to
be performed in less time than the clock period.
Space neither plays a role in the logical function.
The whole circuit may be considered as local, i.e.
propagation times need only to be taken into account when
circuits are connected on large distances,
that is when propagation times are large
when compared to the clock period, as is the case when
computers are connected.
Implementation on a silicon chip must take into account and control
all propagation times within a circuit, so that to insure that all
inputs become stable before the end of each clock
period. The clock signal must be implemented so that it arrives simultaneously
at all {\it latches}, that is with negligible delays when compared to the
clock period. Clearly, such properties can only be checked once the whole
circuit has been specified. Such type of circuit cannot be implemented
incrementally, i.e. by implementing a part without any knowledge
on its connections with other parts and on the clock frequency of the whole
circuit. In other words, all parts must be local at every scale,
from primitives to the whole circuit.

The synchronous approach is however questioned in present VLSI designs.
This arises mainly
because of difficulties which are encountered when distributing 
simultaneously a same clock
signal to millions of {\it latches},
over several ${\rm cm}^2$, and at a frequency of the
order of a Gigaherthz. Clock distribution results
in using an important
part of the chip surface and in producing 
an important part of the overall dissipation.

\bigskip
\noindent {\bf 4.2 Asynchronous circuits}

\noindent Asynchronous circuits may be defined 
in opposition to synchronous circuits,
by the requirement of not using a global clock.
But, rather than being complementary, the class they build includes synchronous
circuits.

Usual models represent asynchronous circuits
as general devices which are distributed
and communicate along connecting channels, as shown on Figure~7.
The activity of such circuits is not ruled by the pulses of a global clock,
but proceeds with communications distributed between many
concurrent parts.
These devices can be simple logical gates (a few transistors) or, at
the opposite, complex processors.
Communications can be realized through a single
wire or through a complex network.
Clearly, concurrency cannot be ignored any more. Indeed, one can no more
define a logical state which would be associated with the global circuit
at a definite time.
For each device can change its state following a communication,
without being synchronized with most other devices.
The notion of computing step itself must be revised, as it relies
on a total ordering of all logical events.

\bigskip

\begin{picture}(0,0)%
\psfig{file=async.pstex}%
\end{picture}%
\setlength{\unitlength}{4144sp}%
\begingroup\makeatletter\ifx\SetFigFont\undefined%
\gdef\SetFigFont#1#2#3#4#5{%
  \reset@font\fontsize{#1}{#2pt}%
  \fontfamily{#3}\fontseries{#4}\fontshape{#5}%
  \selectfont}%
\fi\endgroup%
\begin{picture}(2319,1374)(1024,-2008)
\put(2071,-916){\makebox(0,0)[lb]{\smash{\SetFigFont{12}{14.4}{\rmdefault}{\mddefault}{\updefault}D$_{1}$}}}
\put(1081,-1411){\makebox(0,0)[lb]{\smash{\SetFigFont{12}{14.4}{\rmdefault}{\mddefault}{\updefault}$Env.$}}}
\put(2971,-1141){\makebox(0,0)[lb]{\smash{\SetFigFont{12}{14.4}{\rmdefault}{\mddefault}{\updefault}D$_{2}$}}}
\put(2206,-1861){\makebox(0,0)[lb]{\smash{\SetFigFont{12}{14.4}{\rmdefault}{\mddefault}{\updefault}D$_{3}$}}}
\end{picture}

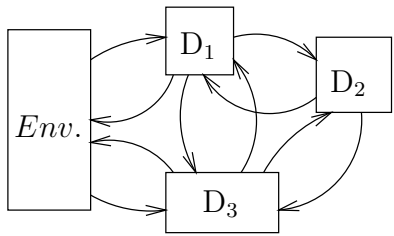
Fig. 7. Asynchronous circuit.

\bigskip
Problems of computation interference may then arise at two different levels.
At lowest level, functioning of a single component may be endangered by 
computing interferences within the component itself, due to internal loops
and instabilities of internal variables. At highest level, composition of
asynchronous circuits may induce computation interferences due to 
exchanges between one module and its environment, the latter 
sending signals which conflict with the module operation.

We shall discuss the second case only and shall assume that primitives
may be defined which are free from internal
computing interferences (see for instance 
\cite{Molnar-Fang-Rosenberger,Furber-Davis}).
We first 
briefly describe those that are most 
frequently used in asynchronous circuits. In some examples, logical functions
are defined in a way which does not distinguish between rising and falling
edges. These undistinguished transitions are called events, and the 
logical function operates on these events. But the events are still
transitions between different levels (or Boolean variables), so that 
each primitive can be considered in both ways, either as an operation
on Boolean variables or as (another) operation on events.

A most frequently used primitive is the {\it join} element, or Muller's
{\it C-element}. It has two inputs $x_1$ and $x_2$ and one output $z$,
and its logical function can be described in the following way:

- if $x_1 = x_2$, then $z = x_1 = x_2$

- if inputs become different from one another, 
then $z$ keeps its previous value.

Then, ouput $z$ only changes after both inputs $x_1$ and $x_2$ have changed. 
This allows a rendez-vous to be realized between levels (wait until
two inputs acquire the same value) or between events (wait until two inputs
 have received the same number of rising or falling edges, after proper
initialization).
 
The primitive {\it C-element} can be realized 
using an electrical feed-back analogous to that of Figure~4.
Along this line, a frequent
realization is represented on Figure~8. 

\bigskip

\begin{picture}(0,0)%
\psfig{file=c-trans.pstex}%
\end{picture}%
\setlength{\unitlength}{4144sp}%
\begingroup\makeatletter\ifx\SetFigFont\undefined%
\gdef\SetFigFont#1#2#3#4#5{%
  \reset@font\fontsize{#1}{#2pt}%
  \fontfamily{#3}\fontseries{#4}\fontshape{#5}%
  \selectfont}%
\fi\endgroup%
\begin{picture}(2352,2004)(451,-1738)
\put(2656,-646){\makebox(0,0)[lb]{\smash{\SetFigFont{12}{14.4}{\rmdefault}{\mddefault}{\updefault}$z$}}}
\put(1711,-1681){\makebox(0,0)[lb]{\smash{\SetFigFont{12}{14.4}{\rmdefault}{\mddefault}{\updefault}0}}}
\put(1711, 74){\makebox(0,0)[lb]{\smash{\SetFigFont{12}{14.4}{\rmdefault}{\mddefault}{\updefault}1}}}
\put(2161,-331){\makebox(0,0)[lb]{\smash{\SetFigFont{10}{12.0}{\rmdefault}{\mddefault}{\updefault}weak}}}
\put(451,-466){\makebox(0,0)[lb]{\smash{\SetFigFont{12}{14.4}{\rmdefault}{\mddefault}{\updefault}$x_{1}$}}}
\put(451,-826){\makebox(0,0)[lb]{\smash{\SetFigFont{12}{14.4}{\rmdefault}{\mddefault}{\updefault}$x_{2}$}}}
\end{picture}

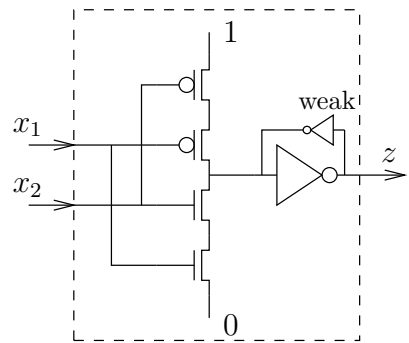
Fig. 8. Simplified C-element.

\bigskip
Another primitive is the {\it toggle}, represented in Figure~9, 
which possesses one input $x$ and two ouputs $z_1$ and $z_2$.
Successive events on the input are alternatively sent to outputs 
$z_1$ and $z_2$. The first event after initialization is sent to
the marked output $z_1$. 

The {\it or} operation between events, also called {\it merge}, can be
realized with a classical {\it exclusive} {\it or} gate 
({\it xor} between levels).
The {\it sequencer}, represented in Figure~9,
possesses three inputs $x_1$, $x_2$ and $x_3$
and two outputs $z_1$ and $z_2$. 
Its role is to grant a given resource to one of two different 
processes which can make requests on inputs $x_1$ and  $x_2$.
When an event is received on $x_3$, a granting event is produced either on
$z_1$ or $z_2$ according to an existing request respectively on $x_1$ or $x_2$.
When two requests are present, the {\it sequencer} arbitrates between the two, 
and thus introduces some part of indeterminism. The {\it sequencer} may take an 
unbounded time to arbitrate, but it is required to realize the mutual 
exclusion of the two grant signals.

\bigskip

\begin{picture}(0,0)%
\psfig{file=prim-di.pstex}%
\end{picture}%
\setlength{\unitlength}{4144sp}%
\begingroup\makeatletter\ifx\SetFigFont\undefined%
\gdef\SetFigFont#1#2#3#4#5{%
  \reset@font\fontsize{#1}{#2pt}%
  \fontfamily{#3}\fontseries{#4}\fontshape{#5}%
  \selectfont}%
\fi\endgroup%
\begin{picture}(1947,3435)(901,-3661)
\put(1891,-1276){\makebox(0,0)[lb]{\smash{\SetFigFont{12}{14.4}{\rmdefault}{\mddefault}{\updefault}$x$}}}
\put(2476,-3661){\makebox(0,0)[lb]{\smash{\SetFigFont{12}{14.4}{\rmdefault}{\mddefault}{\updefault}$x_{3}$}}}
\put(2791,-2131){\makebox(0,0)[lb]{\smash{\SetFigFont{12}{14.4}{\rmdefault}{\mddefault}{\updefault}$z$}}}
\put(2791,-1231){\makebox(0,0)[lb]{\smash{\SetFigFont{12}{14.4}{\rmdefault}{\mddefault}{\updefault}$z_{1}$}}}
\put(2746,-2806){\makebox(0,0)[lb]{\smash{\SetFigFont{12}{14.4}{\rmdefault}{\mddefault}{\updefault}$z_{1}$}}}
\put(2791,-1456){\makebox(0,0)[lb]{\smash{\SetFigFont{12}{14.4}{\rmdefault}{\mddefault}{\updefault}$z_{2}$}}}
\put(2746,-3076){\makebox(0,0)[lb]{\smash{\SetFigFont{12}{14.4}{\rmdefault}{\mddefault}{\updefault}$z_{2}$}}}
\put(1936,-1951){\makebox(0,0)[lb]{\smash{\SetFigFont{12}{14.4}{\rmdefault}{\mddefault}{\updefault}$x_{1}$}}}
\put(1936,-2176){\makebox(0,0)[lb]{\smash{\SetFigFont{12}{14.4}{\rmdefault}{\mddefault}{\updefault}$x_{2}$}}}
\put(1936,-2761){\makebox(0,0)[lb]{\smash{\SetFigFont{12}{14.4}{\rmdefault}{\mddefault}{\updefault}$x_{1}$}}}
\put(1936,-3031){\makebox(0,0)[lb]{\smash{\SetFigFont{12}{14.4}{\rmdefault}{\mddefault}{\updefault}$x_{2}$}}}
\put(2341,-691){\makebox(0,0)[lb]{\smash{\SetFigFont{12}{14.4}{\rmdefault}{\mddefault}{\updefault}C}}}
\put(1891,-691){\makebox(0,0)[lb]{\smash{\SetFigFont{12}{14.4}{\rmdefault}{\mddefault}{\updefault}$x_{2}$}}}
\put(1891,-421){\makebox(0,0)[lb]{\smash{\SetFigFont{12}{14.4}{\rmdefault}{\mddefault}{\updefault}$x_{1}$}}}
\put(901,-691){\makebox(0,0)[lb]{\smash{\SetFigFont{12}{14.4}{\rmdefault}{\mddefault}{\updefault}C-element}}}
\put(901,-1411){\makebox(0,0)[lb]{\smash{\SetFigFont{12}{14.4}{\rmdefault}{\mddefault}{\updefault}Toggle}}}
\put(901,-2221){\makebox(0,0)[lb]{\smash{\SetFigFont{12}{14.4}{\rmdefault}{\mddefault}{\updefault}Merge}}}
\put(901,-2986){\makebox(0,0)[lb]{\smash{\SetFigFont{12}{14.4}{\rmdefault}{\mddefault}{\updefault}Sequencer}}}
\put(2791,-601){\makebox(0,0)[lb]{\smash{\SetFigFont{12}{14.4}{\rmdefault}{\mddefault}{\updefault}$z$}}}
\end{picture}

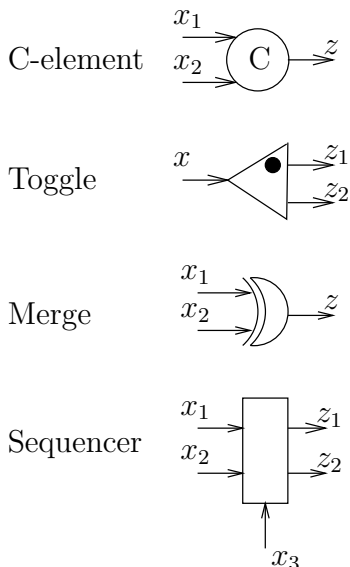
Fig. 9. Some asynchronous primitives.

\section{Composition of circuits and time ordering}

In this section, we discuss the composition of asynchronous circuits,
and some solutions which have been brought to the problem of computing 
interferences. 

Compositions of asynchronous circuits correspond to distributed systems,
where different parts communicate in a way which is 
not regularized by a global clock. 
Then various and arbitrary time delays affect successive transitions 
at the input of a module.
Inputs may then conflict with the correct operation
of the module itself.
The approaches followed in circuit design to deal with
computing interferences fall into two main classes.
One practical approach to timing problems
consists in working directly on the physical 
implementation, by keeping track of all the delays occuring in the logical
circuit, together with all the constraints which must be satisfied by
these delays in order to make the whole circuit operate correctly. 
Then, programs are developed to find and optimize solutions in a
systematic way \cite{Chakraborty-Yun-Dill,Stevens-Ginosar-Rotem}.
Although practically very efficient, this strategy rapidly attains such a
complexity that it becomes very difficult to distinguish fundamental
issues from practical choices. 
In the other class of approaches, one attempts to separate as much as possible
the logical issues related with timing from their physical manifestations,
that is mainly from the values of time delays. 
This has led to different studies, focussing either on the determination
of a best choice of logical primitives, satisfying criteria like
speed-insensitivity or delay-insensitivity \cite{Martin-Hoare},
 or on a more restrictive
definition of modular composition, like delay-insensitive compositions
\cite{Udding-DI-def,Ebergen-Dist.Comp.}. 
In the following, we shall only briefly discuss approaches 
related with delay-insensitivity,
and focus on the fundamental relation they tend to exhibit between the
 occurence of computing interferences and the causal structure of physical
space-time.

In order to make the analysis easier to follow, we shall introduce a graphic
representation of the communications occuring between modules of a 
composition (see Figure~11). 
These graphs are analogous to those that can be used
in relativistic physics to represent the space-time evolution of
localized physical systems, together with the light signals they exchange.
As discussed in a previous section,
an essential feature is the absence of an a priori given
  global and common time.
Only a local time ordering can be made between the successive events occuring 
on each module, reflecting the causal relations which can be made locally.
Although time is represented as the vertical axis, 
this only indicates the direction for 
increasing time on each module.
Different modules are displayed on the horizontal axis, which
roughly corresponds to space.
Each module is then represented by a vertical line, indicating the causal
succession of the local events occuring at its inputs or outputs.
Communications are then represented by inclined arrows leaving  
a module (output) to reach another module (input).
Although they may vary, the slopes of theses arrows must always be greater 
than a strictly positive lower bound, which corresponds to light velocity. 
Varying slopes indicate that varying speed and
delays affect communications between modules.

In the following, we shall denote by "event" each arrow corresponding to a
communication, and shall call "point" the intersection of this event
with the time evolution of a module (following in that way the
notation
introduced by Russell in his discussion of the causal structure
of relativistic space-time \cite{Russell-AM}).
The logical specification of each module is translated into causal relations 
between the points which represent the occurence of events on the module.
These local constraints may be given a precise expression 
using a formal language
well suited to represent time
ordered event structures \cite{Martin-Hoare,Ebergen-Dist.Comp.,Winskel}.
As propagation delays play an essential part,
ordering constraints will be most conveniently visualized on
graphic representations, which allow the analysis of global
causal relations within distributed systems.

\bigskip
\noindent {\bf 5.1 Delay-insensitivity}

\noindent In order to discuss the role of delays in computation interferences,
 let us first analyse 
the illustrating example of the {\it Q-element}
 \cite{Martin-Hoare,Martin-Limit.},
which is represented in Figure~10.
The formal expression describing the logical function
of the {\it Q-element} can be written in a language which is derived from CSP
(Communicating Sequential Processes) \cite{Hoare-CSP}:
\begin{equation}
\label{Qelement}
\ast \left[ [x_i]; y_o\uparrow; [y_i]; u\uparrow; [u]; y_o\downarrow;
[\lnot y_i]; x_o\uparrow; [\lnot x_i]; u\downarrow; [\lnot u]; x_o\downarrow
\right] 
\end{equation}
Each variable between brackets, which precedes
a transition, represents a logical
variable which must be true before the circuit can execute the  
transition which follows ($;$ denotes time succession, and $*$ arbitrary 
repetition of the expresssion in brackets).
Thus, the circuit waits for $x_i$ to be true, then
emits a rising edge on output $y_o$, etc...

This logical function can be implemented as a composition of a {\it C-element}
with two {\it and } gates, as represented in Figure~10. Output $u$
of the {\it C-element} is followed by a fork, which relates $u$ to one input
of each of the {\it and } gates. Two other forks also dispatch
the event produced by the environment $x_i$ (resp. $y_i$)
 on two inputs
denoted by $x_1$ and $x_2$ (resp. $y_1$ and $y_2$).

\bigskip

\begin{picture}(0,0)%
\psfig{file=q-elt.pstex}%
\end{picture}%
\setlength{\unitlength}{4144sp}%
\begingroup\makeatletter\ifx\SetFigFont\undefined%
\gdef\SetFigFont#1#2#3#4#5{%
  \reset@font\fontsize{#1}{#2pt}%
  \fontfamily{#3}\fontseries{#4}\fontshape{#5}%
  \selectfont}%
\fi\endgroup%
\begin{picture}(2712,1872)(1216,-2773)
\put(2251,-2581){\makebox(0,0)[lb]{\smash{\SetFigFont{12}{14.4}{\rmdefault}{\mddefault}{\updefault}$y_{1}$}}}
\put(2476,-1996){\makebox(0,0)[lb]{\smash{\SetFigFont{12}{14.4}{\rmdefault}{\mddefault}{\updefault}C}}}
\put(3826,-1186){\makebox(0,0)[lb]{\smash{\SetFigFont{12}{14.4}{\rmdefault}{\mddefault}{\updefault}$y_{o}$}}}
\put(3016,-1096){\makebox(0,0)[lb]{\smash{\SetFigFont{12}{14.4}{\rmdefault}{\mddefault}{\updefault}$x_{1}$}}}
\put(2251,-1591){\makebox(0,0)[lb]{\smash{\SetFigFont{12}{14.4}{\rmdefault}{\mddefault}{\updefault}$x_{2}$}}}
\put(3061,-1546){\makebox(0,0)[lb]{\smash{\SetFigFont{12}{14.4}{\rmdefault}{\mddefault}{\updefault}$u$}}}
\put(3736,-2581){\makebox(0,0)[lb]{\smash{\SetFigFont{12}{14.4}{\rmdefault}{\mddefault}{\updefault}$y_{i}$}}}
\put(1216,-1096){\makebox(0,0)[lb]{\smash{\SetFigFont{12}{14.4}{\rmdefault}{\mddefault}{\updefault}$x_{i}$}}}
\put(1216,-2491){\makebox(0,0)[lb]{\smash{\SetFigFont{12}{14.4}{\rmdefault}{\mddefault}{\updefault}$x_{o}$}}}
\put(1621,-2626){\makebox(0,0)[lb]{\smash{\SetFigFont{12}{14.4}{\rmdefault}{\mddefault}{\updefault}B}}}
\put(3421,-1321){\makebox(0,0)[lb]{\smash{\SetFigFont{12}{14.4}{\rmdefault}{\mddefault}{\updefault}A}}}
\put(2026,-2266){\makebox(0,0)[lb]{\smash{\SetFigFont{12}{14.4}{\rmdefault}{\mddefault}{\updefault}$u$}}}
\put(2926,-2221){\makebox(0,0)[lb]{\smash{\SetFigFont{12}{14.4}{\rmdefault}{\mddefault}{\updefault}$y_{2}$}}}
\end{picture}

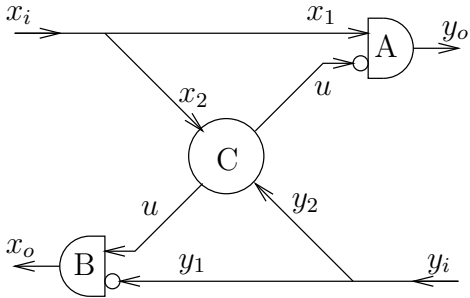
Fig. 10. Q-element.        

\bigskip
The logical operation of the circuit may be represented using a space-time
graph, as in Figure~11. Left and right parts of the circuit environment are
respectively represented as $X$ and $Y$.
The series of points corresponding to the definition of the logical
function of each module can be followed on each vertical line.
Situations which correspond to rendez-vous, i.e. intervals
where a primitive is waiting for the arrival of two events in any order,
have been represented by a thick line.
This is systematically the case for the {\it C-element}, but also for
the {\it and} gates, when they are waiting for their two inputs to be true.
Pairs of points which cannot occur in reverse order without
ruining computation, have been signalized by dashed lines.
The two cases involve the
internal variable $u$ and one event, $y_1\uparrow$ or $x_1\downarrow$
 produced by the
environment ($Y$ or $X$). Event $y_1\uparrow$ must reach the {\it and} gate B
before event $u\uparrow$,
recalling that the latter has been produced by the arrival
of event $y_2\uparrow$ on the {\it C-element}. Then, the fork which dispatches
both events $y_1\uparrow$ and $y_2\uparrow$
plays a crucial role in determining the order
of points on {\it and} gate B.

A few remarks are in order. Concurrent computing is well illustrated
by Figure~11. Different computations proceed along paths involving
vertical and propagation lines,
each representing a causally ordered series of operations.
Causal order makes only sense either within each vertical line, where it
is associated with the logical function of the module, or within propagating
lines, where it connects the output of one module to
the input of another module. But no a priori total order exists between all
points of the graph.
This is illustrated by the independence of
computation on the order of some pairs of points. For instance,
two events belonging to different branches of the fork
on variable $u$ at the output of the {\it C-element}
may have arbitrary relative order.

\bigskip

\begin{picture}(0,0)%
\psfig{file=q-esp-tps.pstex}%
\end{picture}%
\setlength{\unitlength}{4144sp}%
\begingroup\makeatletter\ifx\SetFigFont\undefined%
\gdef\SetFigFont#1#2#3#4#5{%
  \reset@font\fontsize{#1}{#2pt}%
  \fontfamily{#3}\fontseries{#4}\fontshape{#5}%
  \selectfont}%
\fi\endgroup%
\begin{picture}(2757,5487)(1384,-6418)
\put(2071,-5416){\makebox(0,0)[lb]{\smash{\SetFigFont{12}{14.4}{\rmdefault}{\mddefault}{\updefault}$x_{2}\uparrow$}}}
\put(1891,-6181){\makebox(0,0)[lb]{\smash{\SetFigFont{12}{14.4}{\rmdefault}{\mddefault}{\updefault}$X$}}}
\put(2971,-6181){\makebox(0,0)[lb]{\smash{\SetFigFont{12}{14.4}{\rmdefault}{\mddefault}{\updefault}C}}}
\put(4051,-6181){\makebox(0,0)[lb]{\smash{\SetFigFont{12}{14.4}{\rmdefault}{\mddefault}{\updefault}$Y$}}}
\put(1441,-1096){\makebox(0,0)[lb]{\smash{\SetFigFont{12}{14.4}{\rmdefault}{\mddefault}{\updefault}$t$}}}
\put(2431,-6181){\makebox(0,0)[lb]{\smash{\SetFigFont{12}{14.4}{\rmdefault}{\mddefault}{\updefault}B}}}
\put(3511,-6181){\makebox(0,0)[lb]{\smash{\SetFigFont{12}{14.4}{\rmdefault}{\mddefault}{\updefault}A}}}
\put(2026,-5956){\makebox(0,0)[lb]{\smash{\SetFigFont{12}{14.4}{\rmdefault}{\mddefault}{\updefault}$x_{o}\downarrow$}}}
\put(3241,-5956){\makebox(0,0)[lb]{\smash{\SetFigFont{12}{14.4}{\rmdefault}{\mddefault}{\updefault}$u\downarrow$}}}
\put(3151,-5506){\makebox(0,0)[lb]{\smash{\SetFigFont{12}{14.4}{\rmdefault}{\mddefault}{\updefault}$x_{1}\uparrow$}}}
\put(3736,-5281){\makebox(0,0)[lb]{\smash{\SetFigFont{12}{14.4}{\rmdefault}{\mddefault}{\updefault}$y_{o}\uparrow$}}}
\put(4141,-5056){\makebox(0,0)[lb]{\smash{\SetFigFont{12}{14.4}{\rmdefault}{\mddefault}{\updefault}$y_{i}\uparrow$}}}
\put(2566,-4966){\makebox(0,0)[lb]{\smash{\SetFigFont{12}{14.4}{\rmdefault}{\mddefault}{\updefault}$y_{1}\uparrow$}}}
\put(3151,-4381){\makebox(0,0)[lb]{\smash{\SetFigFont{12}{14.4}{\rmdefault}{\mddefault}{\updefault}$u\uparrow$}}}
\put(2656,-4381){\makebox(0,0)[lb]{\smash{\SetFigFont{12}{14.4}{\rmdefault}{\mddefault}{\updefault}$u\uparrow$}}}
\put(3736,-4291){\makebox(0,0)[lb]{\smash{\SetFigFont{12}{14.4}{\rmdefault}{\mddefault}{\updefault}$y_{o}\downarrow$}}}
\put(2566,-3796){\makebox(0,0)[lb]{\smash{\SetFigFont{12}{14.4}{\rmdefault}{\mddefault}{\updefault}$y_{1}\downarrow$}}}
\put(1621,-3211){\makebox(0,0)[lb]{\smash{\SetFigFont{12}{14.4}{\rmdefault}{\mddefault}{\updefault}$x_{i}\downarrow$}}}
\put(3106,-2491){\makebox(0,0)[lb]{\smash{\SetFigFont{12}{14.4}{\rmdefault}{\mddefault}{\updefault}$u\downarrow$}}}
\put(2656,-2491){\makebox(0,0)[lb]{\smash{\SetFigFont{12}{14.4}{\rmdefault}{\mddefault}{\updefault}$u\downarrow$}}}
\put(2026,-2356){\makebox(0,0)[lb]{\smash{\SetFigFont{12}{14.4}{\rmdefault}{\mddefault}{\updefault}$x_{o}\downarrow$}}}
\put(1621,-2041){\makebox(0,0)[lb]{\smash{\SetFigFont{12}{14.4}{\rmdefault}{\mddefault}{\updefault}$x_{i}\uparrow$}}}
\put(1621,-5731){\makebox(0,0)[lb]{\smash{\SetFigFont{12}{14.4}{\rmdefault}{\mddefault}{\updefault}$x_{i}\uparrow$}}}
\put(2116,-2896){\makebox(0,0)[lb]{\smash{\SetFigFont{12}{14.4}{\rmdefault}{\mddefault}{\updefault}$x_{2}\downarrow$}}}
\put(3106,-3031){\makebox(0,0)[lb]{\smash{\SetFigFont{12}{14.4}{\rmdefault}{\mddefault}{\updefault}$x_{1}\downarrow$}}}
\put(2026,-3481){\makebox(0,0)[lb]{\smash{\SetFigFont{12}{14.4}{\rmdefault}{\mddefault}{\updefault}$x_{o}\uparrow$}}}
\put(3196,-3526){\makebox(0,0)[lb]{\smash{\SetFigFont{12}{14.4}{\rmdefault}{\mddefault}{\updefault}$y_{2}\downarrow$}}}
\put(4141,-3976){\makebox(0,0)[lb]{\smash{\SetFigFont{12}{14.4}{\rmdefault}{\mddefault}{\updefault}$y_{i}\downarrow$}}}
\put(3691,-4786){\makebox(0,0)[lb]{\smash{\SetFigFont{12}{14.4}{\rmdefault}{\mddefault}{\updefault}$y_{2}\uparrow$}}}
\end{picture}

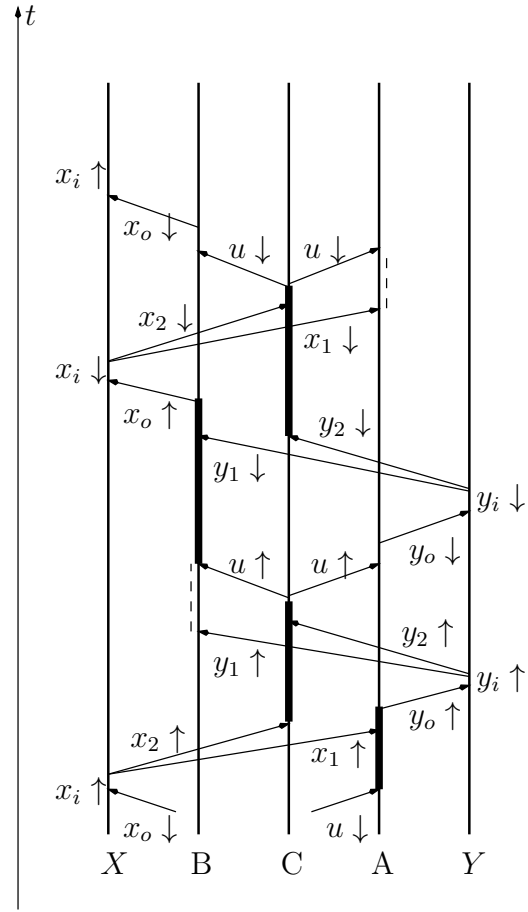
Fig. 11. Space-time graph of the Q-element.

\bigskip
Imposing a total ordering would amount to implement a global time,
by means of clock distribution for instance, which would allow one to draw
horizontal lines on the graph of Figure~11. But such condition is too
restrictive, as computation only relies on causal relations imposed
by vertical and propagation lines. The remaining freedom in the ordering of
events, as the one
related to the fork at the output of the {\it C-element},
is necessary for  optimizing the circuit performance.
For a definite implementation, event ordering will depend on the
relative spatial localization of modules, so that the remaining
freedom may be
used to find an optimal arrangement of modules on the chip.

The property of delay-insensitivity \cite{Molnar-Fang-Rosenberger}
is easily seen on the graph.
It corresponds to the independence of causal ordering of computation
on delays
occuring in responses of modules or in propagations of signals,
i.e. on vertical or horizontal displacements of the modules.
Such property is made possible by using primitives which wait for the
arrival of events at their input before producing other events
at their output. But this condition appears to be unsufficient.
In that respect, it is instructive to compare the two kinds
of forks used by the previous composition implementing the {\it Q-element}.
 No constraint affects the events produced by
the fork at the output
of the {\it C-element} (thick lines in Figure~11).
However, forks dispatching the events produced by the
environment must be implemented in such a way to respect the causal order
of the events which they generate and which finally arrive at the same
{\it and} gate (dashed lines in Figure~11).
Such forks, which are called {\it isochronic forks}
\cite{Martin-Hoare,Martin-Limit.},
must be isolated and given a special treatment
at the implementation level,
in order to satisfy the delay constraints which are necessary for preserving
causal ordering.

The property of delay-insensitivity (DI) has been introduced and much developed
as a simple condition one can impose on primitives and logical
circuits, with the aim to design in a systematical way asynchronous circuits
of arbitrary complexity, without having to take time scales into account.
One approach consists in defining DI circuits as compositions of
stable primitives devoid of internal loops (only electrical loops being
used for memorization) \cite{Martin-Dist.Comp.}.
A primitive is defined to be stable, by imposing that an
input which changes the output cannot change before the output
has been established. It can then be shown that only compositions
of {\it C-elements} can be DI according to this definition.

But, it can also be shown that compositions using {\it C-elements}
(and generalized {\it C-elements} with more inputs) exclusively, strongly
limit the type of allowed computations,
excluding most circuits of interest \cite{Martin-Limit.}.
The {\it isochronic fork} may then be advocated as a weakest compromise
to delay insensitivity. Adding the {\it isochronic fork} and
using this extended class of elements, called quasi
delay-insensitive (QDI),
complex and efficient asynchronous circuits have been realized \cite{R3000}.
However, as illustrated by the example of
the {\it Q-element}, {\it isochronic forks} need to be
identified at the logical level and their implementation
must be given a special treatment, which
may reveal itself intricate for very complex circuits.

\bigskip
\noindent {\bf 5.2 Delay insensitive composition}

\noindent Another approach for avoiding computation
interferences \cite{Molnar-Fang-Rosenberger,Udding-DI-def,Ebergen-Dist.Comp.},
consists in defining a less restrictive set of 
DI primitives, together with a notion of 
DI composition of these primitives.
Circuits are represented in a formal language, called trace theory,
similar to the one used in equation (\ref{Qelement}), with further 
syntax rules on
logical operations. 
Computing interferences are avoided by imposing structural constraints
under the form of simple rules. 
Let us first recall definitions and some properties of trace structures
\cite{Ebergen-Dist.Comp.,Mazurkiewicz}.

\newtheorem{definition}{Definition}
\begin{definition}
Trace structures are defined as triples
$R= <{\bf i}R, {\bf o}R, {\bf t}R>$,
where ${\bf i}R$ and ${\bf o}R$ are finite sets of symbols, respectively the
input alphabet and the output alphabet,
 and ${\bf t}R$ is the set  of traces, which is a subset of
$({\bf i}R \cup {\bf o}R)^*$,
the set of all finite-length sequences of symbols taken in the union
set ${\bf i}R \cup {\bf o}R$.
\end{definition}

Trace structures are traditionally denoted by
capital letters, while lower case letters $a$, $b$, $c$ denote symbols
and $s$, $t$ traces.
The following short notations are also frequently used:
\beqa
&& a? \equiv
 <\lbrace a \rbrace, \emptyset, \lbrace a \rbrace> \nonumber\\
&& b! \equiv
<\emptyset, \lbrace b \rbrace, \lbrace b \rbrace>
\eeqa

\begin{definition}
Operations of concatenation, union, repetition, prefix-closure, projection and
weaving are defined on trace structures:
\beqa
\label{operations}
&&R;S \equiv <{\bf i}R \cup {\bf i}S, {\bf o}R \cup {\bf o}S,
({\bf t}R) ({\bf t}S)> \nonumber\\
&&R|S  \equiv <{\bf i}R \cup {\bf i}S, {\bf o}R \cup {\bf o}S,
{\bf t}R \cup {\bf t}S> \nonumber\\
&&*[R]  \equiv <{\bf i}R, {\bf o}R, ({\bf t}R)^*>  \nonumber\\
&&{\bf pref}R  \equiv <{\bf i}R, {\bf o}R, \lbrace t_0|\exists t_1 : t_0t_1
\in {\bf t}R\rbrace> \nonumber\\
&&R\downarrow A  \equiv
 <{\bf i}R \cap A, {\bf o}R \cap A, \lbrace t\downarrow A|
t \in {\bf t}R \rbrace> \nonumber\\
&&R||S  \equiv <{\bf i}R \cup {\bf i}S, {\bf o}R \cup {\bf o}S, \nonumber\\
&& \qquad \quad
\lbrace t \in ({\bf a}R \cup {\bf a}S)^* | t\downarrow {\bf a}R \in
{\bf t}R \wedge t\downarrow {\bf a}S \in {\bf t}S \rbrace>
\eeqa
\end{definition}
where, for convenience, notation ${\bf a}R \equiv {\bf i}R \cup {\bf o}R$
has been introduced for the total alphabet of $R$, where
$t\downarrow A$ denotes the projection of trace $t$ on alphabet $A$
and $({\bf t}R)^*$ is the set of all finite-length concatenations of
traces in ${\bf t}R$ (symbols $\exists$, $\forall$, $\in$, $\cap$ and $\cup$
 denote as usual,
existence, universality, set belonging, set intersection and set union).
The ${\bf pref}$ operator constructs prefix-closed
structures, while the projection operator hides internal symbols; finally, the
weave operator expresses instantaneous synchronization.
A circuit is specified by a prefix-closed, non empty, trace structure $R$
with ${\bf i}R \cap {\bf o}R = \emptyset$.
The trace structure representing the
environment of a circuit with trace structure $R$ is the
reflection of the latter, and may also be given a compact notation:
\beqn
{\bar R} = <{\bf o}R, {\bf i}R, {\bf t}R>
\eeqn
A trace structure $R$ may be physically
implemented by letting each symbol $a$ in the
alphabets ${\bf i}R$ and ${\bf o}R$ correspond to a channel,
and each occurence of this symbol
in a trace of ${\bf t}R$ correspond to an event, i.e. a high or low
transition, on
the corresponding channel. Symbols in ${\bf i}R$ or ${\bf o}R$ describe
communication actions that are respectively
 produced by the environment or (exclusive or) by the circuit.

In order to be able to ignore transmission delays
while avoiding transmission and computing interferences, the following rules
may be imposed \cite{Udding-DI-def}.

\beqa
&&{\bf R_0} \quad
\forall s\in {\bf t}R, \quad a\in{\bf a}R \qquad s a a
\not\in {\bf t}R \nonumber\\
&&{\bf R_1}  \quad
\forall s,t \in {\bf t}R, \quad (a,b\in{\bf i}R) \vee (a,b\in{\bf o}R)
\nonumber\\
&&\qquad \qquad s a b t \in {\bf t}R \quad \Leftrightarrow \quad
s b a t \in {\bf t}R \nonumber\\
&&{\bf R_2}  \quad
\forall s,t \in {\bf t}R, \quad (a\in{\bf i}R \wedge b\in{\bf o}R)
 \vee (a\in{\bf o}R \wedge b\in{\bf i}R)
\nonumber\\
&& \quad (s a b \in {\bf t}R \wedge s  b a \in {\bf t}R) \quad
\Rightarrow \quad  (s a b t \in {\bf t}R \Leftrightarrow s  b a t \in {\bf t}R)
\eeqa
Rule ${\bf R_0}$ excludes two consecutive transitions on the same wire,
and hence transmission interferences that may result. Rule ${\bf R_1}$
expresses independence of computation on the order of signals travelling
in the same direction, as this order may depend on suffered
delays. 
The {\it C-element} is easily seen to satisfy this rule. However, the {\it and}
gate only complies with the rule when it is waiting for a rising edge on its two
inputs, and does not in all other cases. Thus the {\it and} gate, and also
the {\it or} gate are excluded by this rule, although the {\it toggle},
the {\it merge} and the {\it sequencer} are compatible.
Rule ${\bf R_2}$ expresses the same property
for signals
travelling in opposite directions, in
case their order does not change the result locally.
Note that due to the necessary symmetric treatment of a circuit
and its environment,
all rules are symmetric in the exchange of input and output
symbols.

One must exclude the possibility for a symbol of one type to disable
a symbol of another type (symbol $a$ is said to disable symbol $b$ in trace
structure $R$, if there is a trace $s$ with $sa\in{\bf t}R \wedge sb\in{\bf t}R
\wedge sab\not\in{\bf t}R$). Such exclusion is necessary
to prevent an admissible input symbol
to get disabled by an output signal,
depending on the delay the former has suffered on its way to the
circuit (by symmetry the same property must also hold for the environment
and output signals). Depending on the level of exclusion,
this property leads to define three classes, with
rules ${\bf R_3^\prime}$, ${\bf R_3^{\prime\prime}}$ and
${\bf R_3^{\prime\prime\prime}}$:
\beqa
&&{\bf R_3^\prime} \quad
\forall s, \quad a\not= b\in{\bf a}R  \nonumber\\
&&\qquad (s a \in {\bf t}R \wedge s  b \in {\bf t}R) \quad
\Rightarrow \quad  s a b \in {\bf t}R
\eeqa
\beqa
&&{\bf R_3^{\prime\prime}} \quad
\forall s, \quad a\not= b\in{\bf a}R \quad
a\not\in{\bf i}R \vee b\not\in{\bf i}R \nonumber\\
&&\qquad (s a \in {\bf t}R \wedge s  b \in {\bf t}R) \quad
\Rightarrow \quad  s a b \in {\bf t}R
\eeqa
\beqa
&&{\bf R_3^{\prime\prime\prime}} \quad
\forall s, \quad (a\in{\bf i}R \wedge b\in{\bf o}R)
 \vee (a\in{\bf o}R \wedge b\in{\bf i}R) \nonumber\\
&&\qquad (s a \in {\bf t}R \wedge s  b \in {\bf t}R) \quad
\Rightarrow \quad  s a b \in {\bf t}R
\eeqa
These rules successively allow for more decision possibility.
Rule ${\bf R_3^\prime}$ does not permit data transmission and is called
synchronization class (an example is provided by the {\it C-element}).
Rule ${\bf R_3^{\prime\prime}}$ allows for two inputs to disable each other
and is called data communication class.
With rule ${\bf R_3^{\prime\prime\prime}}$, a circuit may choose 
between two output symbols and belongs to the arbitration class.

Finally, rule ${\bf R_2}$ appears on specific examples to be too restrictive
\cite{Udding-DI-def}.
An alternative and more generally efficient rule is provided
by:
\beqa
&&{\bf R_2^\prime} \quad
\forall s,t \in {\bf t}R, \quad (a, c\in{\bf i}R \wedge b\in{\bf o}R)
 \vee (a, c\in{\bf o}R \wedge b\in{\bf i}R) \nonumber\\
&& \qquad (s a b t c\in {\bf t}R \wedge s  b a t\in {\bf t}R) \quad
\Rightarrow \quad  s b a t c \in {\bf t}R
\eeqa
This rule, which
is conveniently expressed on a space-time graph, as shown in Figure~12,
concerns three events $a$, $b$, $c$
connecting one module M
and its environment $E$.
It stipulates that, if two time orders are allowed for the occurences
of two events of different types (i.e. one input and one output) $a$ and $b$,
then if the event $c$, of the same type as $a$, is a consequence of the
order "$a$ then $b$", it should also be a consequence of the other order
"$b$ then $a$". This rule imposes that if an order on events is differently
seen by a module and its environment, due to propagation time delays,
then this order should have no consequence on the logical behavior
of the module. As illustrated by Figure~12,
this rule only affects the case on the left
part of the figure, that is, only the case when propagation can
change the order of events.

\bigskip

\begin{picture}(0,0)%
\psfig{file=rp2.pstex}%
\end{picture}%
\setlength{\unitlength}{4144sp}%
\begingroup\makeatletter\ifx\SetFigFont\undefined%
\gdef\SetFigFont#1#2#3#4#5{%
  \reset@font\fontsize{#1}{#2pt}%
  \fontfamily{#3}\fontseries{#4}\fontshape{#5}%
  \selectfont}%
\fi\endgroup%
\begin{picture}(2104,1932)(1159,-2368)
\put(1216,-601){\makebox(0,0)[lb]{\smash{\SetFigFont{12}{14.4}{\rmdefault}{\mddefault}{\updefault}$t$}}}
\put(1576,-2311){\makebox(0,0)[lb]{\smash{\SetFigFont{12}{14.4}{\rmdefault}{\mddefault}{\updefault}M}}}
\put(2071,-2311){\makebox(0,0)[lb]{\smash{\SetFigFont{12}{14.4}{\rmdefault}{\mddefault}{\updefault}$E$}}}
\put(2701,-2311){\makebox(0,0)[lb]{\smash{\SetFigFont{12}{14.4}{\rmdefault}{\mddefault}{\updefault}M'}}}
\put(3196,-2311){\makebox(0,0)[lb]{\smash{\SetFigFont{12}{14.4}{\rmdefault}{\mddefault}{\updefault}$E$'}}}
\put(1756,-1141){\makebox(0,0)[lb]{\smash{\SetFigFont{12}{14.4}{\rmdefault}{\mddefault}{\updefault}$c$}}}
\put(2881,-1051){\makebox(0,0)[lb]{\smash{\SetFigFont{12}{14.4}{\rmdefault}{\mddefault}{\updefault}$c$}}}
\put(1756,-1951){\makebox(0,0)[lb]{\smash{\SetFigFont{12}{14.4}{\rmdefault}{\mddefault}{\updefault}$a$}}}
\put(1756,-1501){\makebox(0,0)[lb]{\smash{\SetFigFont{12}{14.4}{\rmdefault}{\mddefault}{\updefault}$b$}}}
\put(2881,-1906){\makebox(0,0)[lb]{\smash{\SetFigFont{12}{14.4}{\rmdefault}{\mddefault}{\updefault}$b$}}}
\put(2881,-1546){\makebox(0,0)[lb]{\smash{\SetFigFont{12}{14.4}{\rmdefault}{\mddefault}{\updefault}$a$}}}
\end{picture}

\bigskip
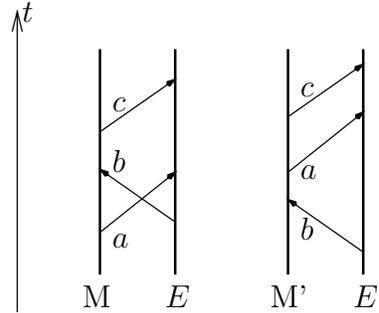
Fig. 12. Space-time graph for rule $R^\prime_2$. 

\bigskip
The set of DI components is given by trace structures, 
defined according to Definitions 1 and 2,
which satisfy the weakest form of the rules, i.e. 
$R_0$, $R_1$, $R_2^\prime$ and $R_3^{\prime\prime\prime}$ \cite{Udding-DI-def}.

A set of DI primitive components for asynchronous circuits
can thus be obtained with the following list
of specifications in terms of trace
structures (see Table~1).

\begin{table}
 \caption{DI primitive components}
 \begin{tabular}{cc}
 \hline \hline
   CIRCUIT & Specification\\
 \hline
 WIRE & ${\bf pref}*[a?;b!]$ \\
& \\
 IWIRE & ${\bf pref}*[b!;a?]$ \\
& \\
 FORK & ${\bf pref}*[a?;b!||c!]$ \\
& \\
 C-ELEMENT & ${\bf pref}*[a?||b?;c!]$ \\
& \\
 TOGGLE & ${\bf pref}*[a?;b!;a?;c!]$\\
& \\
 MERGE & ${\bf pref}*[(a?|b?);c!]$ \\
& \\
 SEQUENCER  & \hskip2mm ${\bf pref}*[a?;p!]$\\
 & $|| {\bf pref}*[b?;q!]$\\
 & \hskip6mm $|| {\bf pref}*[n?;(p!|q!)]$ \\
  \hline \hline
 \end{tabular}
\label{table1}
\end{table}

The {\it wire} corresponds to a component which waits for an event to occur
on its input, then sends an event on its output, and repeats this sequence
indefinitely. The inverted wire ({\it iwire})
behaves similarly, but begins by sending an
event on its output.
The {\it fork} duplicates one input.
As can be seen from definitions (\ref{operations}),
weaving not only consists in putting in parallel, but also in
synchronizing common output symbols. In the particular case
of two {\it wires} with a common output, weaving  leads to the {\it C-element}.
The other components
correspond to the primitive circuits which have been previously introduced
(see Figure~9).

The objective is to realize circuits corresponding to given
complex specifications by combining simple DI primitive circuits.
This aim may be attained by making use of operations such as
decomposition and substitution,
together with two theorems setting the conditions for performing these
operations.

\begin{definition}
A component $R_0$ is said to be decomposed into components
$R_i, 1\leq i< n$ (for $n>1)$ if the following conditions are satisfied.
Letting
$S_0 = {\bar R_0}, \quad S_i = R_i \quad {\rm for}
\quad 1\leq i<n, \quad {\rm and}
\quad T = ||_{0\leq i<n} (S_i)$
\beqa
\label{decomposition}
&&(i) \cup_{0\leq i<n} ({\bf o}S_i) = \cup_{0\leq i<n} ({\bf i}S_i)\nonumber\\
&&(ii) {\bf o}S_i \cap {\bf o}S_j = \emptyset, \quad {\rm for} \quad
0\leq i,j<n \quad {\rm and} \quad
i\not= j\nonumber\\
&&(iii) \forall t,x,i \quad (0\leq i<n)\nonumber\\
&&t\in {\bf t}T \wedge x\in {\bf o}S_i \wedge tx\downarrow {\bf a}S_i
\in {\bf t}S_i \Rightarrow tx\in {\bf t}T \nonumber\\
&&(iv){\bf t}T\downarrow{\bf a}S_0 = {\bf t}S_0
\eeqa
\end{definition}
Conditions in (\ref{decomposition}) respectively describe a closed network (each
 input is connected to an output and conversely (i)),
absence of output interferences (two outputs cannot be connected (ii)),
absence of computing interferences (any event produced by a component is
compatible with the behavior of the component which receives it (iii))
and correct behavior at circuit boundary (network behaves as prescribed (iv)).
Decomposition will be denoted by $R_0 \rightarrow (R_i)_{1\leq i<n}$.

Let us now state two useful theorems (proofs may be obtained in
 \cite{Ebergen-thesis}).

\newtheorem{theorem}{Substitution Theorem}
\begin{theorem}
For components $R_0$, $R_1$, $R_2$, $R_3$ and $S$
\beqa
&&R_0 \rightarrow (R_1, S) \quad \wedge \quad S
\rightarrow (R_2, R_3) \nonumber\\
&& \Rightarrow \qquad R_0 \rightarrow (R_1, R_2, R_3)\nonumber
\eeqa
holds if
\beqn
({\bf a}R_0 \cup {\bf a}R_1) \cap
({\bf a}R_2 \cup {\bf a}R_3) =
{\bf a}S
\eeqn
\end{theorem}
The latter condition stipulates that internal symbols of $S$, i.e. symbols in
$({\bf a}R_2 \cup {\bf a}R_3) \backslash S$, where $\backslash$ means set
deletion, should not appear in $({\bf a}R_0 \cup {\bf a}R_1)$.
It can be realized by an appropriate renaming of internal symbols of $S$.

\newtheorem{stheorem}[theorem]{Separation Theorem}
\begin{stheorem}
For components $R_i$ and $S_i$ ($0\leq i<n$)
\beqa
&&R_0 \rightarrow (R_i)_{1\leq i<n} \quad \wedge \quad
S_0 \rightarrow (S_i)_{1\leq i<n} \nonumber\\
&& \Rightarrow \qquad R_0||S_0 \rightarrow (R_i||S_i)_{1\leq i<n}
\nonumber
\eeqa
holds if
\beqa
&& (\cup_{1\leq i<n}({\bf a}R_i) \backslash {\bf a}R_0)
\cap (\cup_{1\leq i<n}({\bf a}S_i) \backslash {\bf a}S_0)
=\emptyset\label{sep1}\\
&&\quad {\rm and}, \quad \quad {\rm for} \quad 1\leq i\not= j <n\nonumber\\
&&({\bf o}R_i \cup {\bf o}S_i) \cap  ({\bf o}R_j \cup {\bf o}S_j) =\emptyset
\nonumber\\
&&({\bf o}R_i \cup {\bf o}S_i) \cap ({\bf o}{\bar R_0} \cup {\bf o}{\bar S_0})
 =\emptyset\label{sep2}
\eeqa
\end{stheorem}
Condition (\ref{sep1}) stipulates that the internal symbols of
the decompositions
of $R_0$ and $S_0$ are disjoint (this condition may be satisfied by renaming
some of these symbols), and conditions (\ref{sep2}) stipulate
that the outputs of any two components
$R_i||S_i$ and $R_j||S_j$
are also disjoint when the components are different (these conditions may also
be satisfied by reordering the components).

With the help of these two theorems, the previously defined
DI primitives may be combined to give
modular compositions which are delay insensitive, 
hence circuits where computing interferences cannot be introduced by
delay modifications only.
We briefly describe an example of a circuit which can be obtained
with such a composition of DI primitives,
the {\it token-ring}
 interface \cite{Ebergen-Dist.Comp.}.
The {\it token-ring} interface is a device allowing to connect several machines,
which must share a common resource (like a memory, or a bus).
One item ${\rm{Alloc}}_i$
of this device will be associated with each machine $M_i$, all items being
identical and realizing the same function, as shown by Figure~13.
Item ${\rm{Alloc}}_i$ of this device is connected to two environments,
the machine $M_i$ on
top of the figure, and, at bottom, the ring $R$ where a token circulates.
The arrival of the token at ${\rm{Alloc}}_i$ corresponds to an event on $b$,
its departure to an event on $q$.
The machine $M_i$ can make a request under the form of an event on $a_1$.
${\rm{Alloc}}_i$ grants the resource to the machine $M_i$ by an event on $p_1$.
The machine $M_i$ signals the end of its use of the resource by an event on
 $a_0$, which is acknowledged by ${\rm{Alloc}}_i$
under the form of an event on $p_0$.

Initially, the {\it token-ring} interface is specified by the following
 trace structure:
\beqa
&& \hskip2mm {\bf pref}*[a1?;p1!;a0?;p0!]\nonumber\\
&& || {\bf pref}*[b?;(q!|p1!;a0?;q!)]
\eeqa
This specification results from weaving two trace structures which
respectively describe the communications of the token-ring interface with
the machine $M_i$ and with the ring $R$.
 The two trace stuctures interact through
their common dependence on two events $p1$ and $a0$. Each trace structure
may be decomposed into primitive elements. Substitution and separation
theorems may be applied, finally leading to a possible decomposition,  
as shown by Figure~13 in a graphic way:
\beqa
\rightarrow \quad ( && \hskip2mm {\bf pref}*[a1?;p1!]\nonumber\\  
&& || {\bf pref}*[rq1?;q1!]\nonumber\\
&& || {\bf pref}*[b?;(q1!|p1!)],\nonumber
\eeqa
$${\bf pref}*[rq1!;q1?],$$  
$${\bf pref}*[a0?;p0!],$$ 
$${\bf pref}*[a0?;q0!],$$
\beqn
{\bf pref}*[(q1?|q0?);q!)] \quad )
\label{token-dec}
\eeqn
The first component is a {\it sequencer} (see Table~1), necessary for
synchronizing the output $p1$ shared by the two trace structures
defining the {\it token-ring}. 
The {\it sequencer} also arbitrates between corresponding
inputs. Other components describe an {\it iwire}, two {\it wires}
 and a {\it merge}.
Although they do not appear explicitly in decomposition (\ref{token-dec}),
two forks appear in Figure~13, as a consequence of double occurences of 
$a0?$ and $q1?$ in (\ref{token-dec}).

\bigskip

\begin{picture}(0,0)%
\psfig{file=alloc.pstex}%
\end{picture}%
\setlength{\unitlength}{4144sp}%
\begingroup\makeatletter\ifx\SetFigFont\undefined%
\gdef\SetFigFont#1#2#3#4#5{%
  \reset@font\fontsize{#1}{#2pt}%
  \fontfamily{#3}\fontseries{#4}\fontshape{#5}%
  \selectfont}%
\fi\endgroup%
\begin{picture}(2814,2953)(1564,-4169)
\put(3646,-2536){\makebox(0,0)[lb]{\smash{\SetFigFont{10}{12.0}{\rmdefault}{\mddefault}{\updefault}Merge}}}
\put(4186,-1726){\makebox(0,0)[lb]{\smash{\SetFigFont{12}{14.4}{\rmdefault}{\mddefault}{\updefault}$p_{1}$}}}
\put(3241,-2311){\makebox(0,0)[lb]{\smash{\SetFigFont{12}{14.4}{\rmdefault}{\mddefault}{\updefault}D}}}
\put(2341,-2311){\makebox(0,0)[lb]{\smash{\SetFigFont{10}{12.0}{\rmdefault}{\mddefault}{\updefault}Iwire}}}
\put(3151,-2536){\makebox(0,0)[lb]{\smash{\SetFigFont{12}{14.4}{\rmdefault}{\mddefault}{\updefault}$q_{0}$}}}
\put(3556,-1726){\makebox(0,0)[lb]{\smash{\SetFigFont{12}{14.4}{\rmdefault}{\mddefault}{\updefault}$p_{0}$}}}
\put(2161,-2356){\makebox(0,0)[lb]{\smash{\SetFigFont{12}{14.4}{\rmdefault}{\mddefault}{\updefault}A}}}
\put(3691,-2986){\makebox(0,0)[lb]{\smash{\SetFigFont{12}{14.4}{\rmdefault}{\mddefault}{\updefault}C}}}
\put(2341,-2896){\makebox(0,0)[lb]{\smash{\SetFigFont{12}{14.4}{\rmdefault}{\mddefault}{\updefault}B}}}
\put(2521,-3886){\makebox(0,0)[lb]{\smash{\SetFigFont{12}{14.4}{\rmdefault}{\mddefault}{\updefault}$b$}}}
\put(3961,-3886){\makebox(0,0)[lb]{\smash{\SetFigFont{12}{14.4}{\rmdefault}{\mddefault}{\updefault}$q$}}}
\put(2116,-1726){\makebox(0,0)[lb]{\smash{\SetFigFont{12}{14.4}{\rmdefault}{\mddefault}{\updefault}$a_{0}$}}}
\put(1621,-3481){\makebox(0,0)[lb]{\smash{\SetFigFont{12}{14.4}{\rmdefault}{\mddefault}{\updefault}Alloc$_{i}$}}}
\put(2116,-2581){\makebox(0,0)[lb]{\smash{\SetFigFont{12}{14.4}{\rmdefault}{\mddefault}{\updefault}$rq_{1}$}}}
\put(2611,-2761){\makebox(0,0)[lb]{\smash{\SetFigFont{12}{14.4}{\rmdefault}{\mddefault}{\updefault}$q_{1}$}}}
\put(1576,-1726){\makebox(0,0)[lb]{\smash{\SetFigFont{12}{14.4}{\rmdefault}{\mddefault}{\updefault}$a_{1}$}}}
\put(2611,-3301){\makebox(0,0)[lb]{\smash{\SetFigFont{10}{12.0}{\rmdefault}{\mddefault}{\updefault}Sequencer}}}
\put(2926,-4111){\makebox(0,0)[lb]{\smash{\SetFigFont{12}{14.4}{\rmdefault}{\mddefault}{\updefault}Ring ($R$)}}}
\put(2386,-1411){\makebox(0,0)[lb]{\smash{\SetFigFont{12}{14.4}{\rmdefault}{\mddefault}{\updefault}Machine $i$ ($M_{i}$)}}}
\end{picture}

Fig. 13. Token-ring interface.

\bigskip
The DI property of this implementation can be visualized 
on a space-time graph, as in Figure~14.
Two cases have been represented in Figure~14. In the first case, the request
$a_1$ done by the machine $i$ is not granted, the token being sent
back to the ring. When the token arrives a second time, the resource is
granted to the machine $M_i$ which was waiting. This illustrates the
undeterministic behavior of the module $\rm{Alloc}$,
which depends on arbitration
performed by {\it sequencer} B.
The figure also shows that the two forks, that on $q_1$ (output of B)
and that defined by D cannot create computation interferences,
so that no particular constraints are necessary. This results from the
function of {\it sequencer} B, which is not perturbed whatever the order of
the events on its inputs. {\it Sequencer} B waits for an event on $b$ to make a
decision, and then arbitrates between the different requests it has received.

As shown by the example of the {\it token-ring} interface,
DI primitives and DI decomposition may be used to generate
modular compositions which are delay insensitive, and, 
as shown with the help of space-time graphs, that
remain free of computing interferences. 
Delay-insensitivity appears as a simple criterion for 
escaping problems raised by computing interferences in a purely logically
way, i.e. without recourse to a detailed analysis of the 
physical implementation of a circuit. The DI criterion allows one
to treat asynchronous
circuits efficiently, 
like in the case of synchronous circuits,
by allowing to represent them formally 
(in terms of trace structures).
Although revealing a genuinely different underlying structure,
the causal constraints on asynchronous circuits, 
as exhibited by space-time graphs, can nonetheless be embedded
in a simple set of formal rules which limit the definition and composition
of DI circuits.
In general, these rules allow
DI circuits to be decomposed into a number of DI primitive components which
increases linearly with the length of the circuit specification
\cite{Ebergen-thesis}. 

\bigskip

\begin{picture}(0,0)%
\psfig{file=alloc-esp-tps.pstex}%
\end{picture}%
\setlength{\unitlength}{4144sp}%
\begingroup\makeatletter\ifx\SetFigFont\undefined%
\gdef\SetFigFont#1#2#3#4#5{%
  \reset@font\fontsize{#1}{#2pt}%
  \fontfamily{#3}\fontseries{#4}\fontshape{#5}%
  \selectfont}%
\fi\endgroup%
\begin{picture}(3082,5532)(1306,-6418)
\put(2071,-4606){\makebox(0,0)[lb]{\smash{\SetFigFont{10}{12.0}{\rmdefault}{\mddefault}{\updefault}not-granted}}}
\put(3421,-2086){\makebox(0,0)[lb]{\smash{\SetFigFont{12}{14.4}{\rmdefault}{\mddefault}{\updefault}$b$}}}
\put(2296,-2401){\makebox(0,0)[lb]{\smash{\SetFigFont{12}{14.4}{\rmdefault}{\mddefault}{\updefault}$a_{1}$}}}
\put(4051,-2761){\makebox(0,0)[lb]{\smash{\SetFigFont{12}{14.4}{\rmdefault}{\mddefault}{\updefault}$q$}}}
\put(2296,-3436){\makebox(0,0)[lb]{\smash{\SetFigFont{12}{14.4}{\rmdefault}{\mddefault}{\updefault}$p_{1}$}}}
\put(1756,-4381){\makebox(0,0)[lb]{\smash{\SetFigFont{12}{14.4}{\rmdefault}{\mddefault}{\updefault}$rq_{1}$}}}
\put(3421,-5281){\makebox(0,0)[lb]{\smash{\SetFigFont{12}{14.4}{\rmdefault}{\mddefault}{\updefault}$b$}}}
\put(1756,-5281){\makebox(0,0)[lb]{\smash{\SetFigFont{12}{14.4}{\rmdefault}{\mddefault}{\updefault}$rq_{1}$}}}
\put(1621,-6181){\makebox(0,0)[lb]{\smash{\SetFigFont{12}{14.4}{\rmdefault}{\mddefault}{\updefault}A}}}
\put(2161,-6181){\makebox(0,0)[lb]{\smash{\SetFigFont{12}{14.4}{\rmdefault}{\mddefault}{\updefault}B}}}
\put(3241,-6181){\makebox(0,0)[lb]{\smash{\SetFigFont{12}{14.4}{\rmdefault}{\mddefault}{\updefault}D}}}
\put(3781,-6181){\makebox(0,0)[lb]{\smash{\SetFigFont{12}{14.4}{\rmdefault}{\mddefault}{\updefault}C}}}
\put(4321,-6181){\makebox(0,0)[lb]{\smash{\SetFigFont{12}{14.4}{\rmdefault}{\mddefault}{\updefault}$R$}}}
\put(2656,-6181){\makebox(0,0)[lb]{\smash{\SetFigFont{12}{14.4}{\rmdefault}{\mddefault}{\updefault}$M_{i}$}}}
\put(1486,-1051){\makebox(0,0)[lb]{\smash{\SetFigFont{12}{14.4}{\rmdefault}{\mddefault}{\updefault}$t$}}}
\put(2926,-2761){\makebox(0,0)[lb]{\smash{\SetFigFont{12}{14.4}{\rmdefault}{\mddefault}{\updefault}$p_{0}$}}}
\put(3466,-2806){\makebox(0,0)[lb]{\smash{\SetFigFont{12}{14.4}{\rmdefault}{\mddefault}{\updefault}$q_{0}$}}}
\put(2971,-3391){\makebox(0,0)[lb]{\smash{\SetFigFont{12}{14.4}{\rmdefault}{\mddefault}{\updefault}$a_{0}$}}}
\put(3421,-3931){\makebox(0,0)[lb]{\smash{\SetFigFont{12}{14.4}{\rmdefault}{\mddefault}{\updefault}$b$}}}
\put(4006,-4561){\makebox(0,0)[lb]{\smash{\SetFigFont{12}{14.4}{\rmdefault}{\mddefault}{\updefault}$q$}}}
\put(1891,-4741){\makebox(0,0)[lb]{\smash{\SetFigFont{12}{14.4}{\rmdefault}{\mddefault}{\updefault}$q_{1}$}}}
\put(2341,-5731){\makebox(0,0)[lb]{\smash{\SetFigFont{12}{14.4}{\rmdefault}{\mddefault}{\updefault}$a_{1}$}}}
\put(3376,-4561){\makebox(0,0)[lb]{\smash{\SetFigFont{12}{14.4}{\rmdefault}{\mddefault}{\updefault}$q_{1}$}}}
\put(1606,-3706){\makebox(0,0)[lb]{\smash{\SetFigFont{10}{12.0}{\rmdefault}{\mddefault}{\updefault}granted}}}
\end{picture}

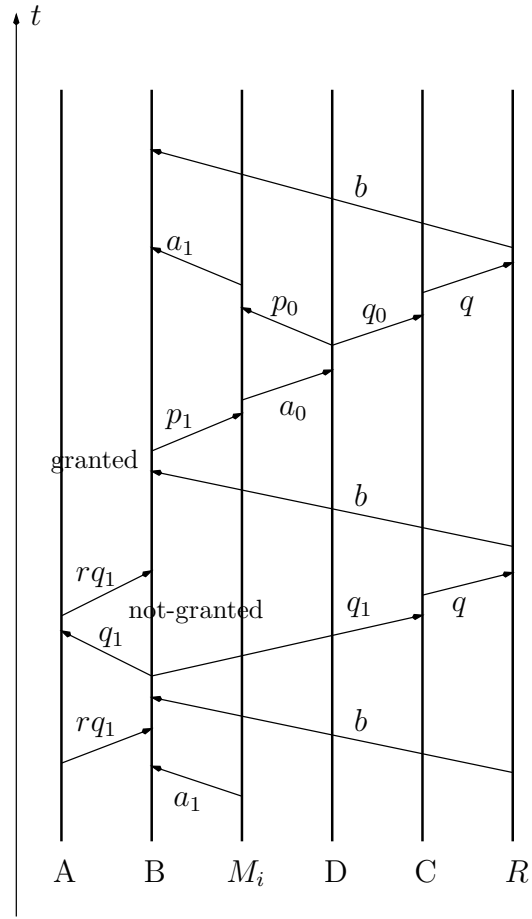
Fig. 14. Space-time graph of the token-ring interface.

\bigskip
\section{Conclusion}

Without giving definite answers to the problems raised in the introduction,
we have nethertheless tried to provide some hints on the essential
role played by
physical time in computation. The necessary reference to physical time
in physical implementations of logical circuits forces one to give
an explicit treatment
of computation interferences. These arise as obstructions
when trying to make
the causality underlying logical circuits coincide 
with the physical causality of their
implementations. For synchronous circuits, these may be avoided by
ruling the whole circuit with a single clock, which thus provides
a global reference to a Newtonian time.
In general however, circuits must be considered as asynchronous and
physical space-time as relativistic.
In the latter, not all points are causally related,
but only those such that one point falls within the light cone
issued from the other. In that respect, asynchronous circuits and 
relativistic space-time share the same founding point of view.
Points derive from events and not the converse, 
propagating events being treated as primary
entities and not as successions of points.
The distinction between two classes of points can also be
seen in a simple way:
two points are causally related if and only if there exists 
a path between them using vertical or propagation lines
(in different spatial directions, 
but in the same time direction); on another hand,
points defined on two different 
events originating from the
same point are not causally related \cite{Russell-AM}.
Similarly for a concurrent computation, 
each computing path connects points which are
causally related. Avoiding
computing interferences corresponds to impose that different computing paths
respect a same
time ordering, but only for pairs of causally related points.

Remedies to computing interferences in asynchronous circuits
consist in recognizing paths which may conflict with a module specification, 
and in eventually delaying these paths, 
so that to respect a prescribed time ordering. 
This can be done either physically, at the implementation level by introducing
explicit time delays, or at the
logical level, by imposing specification rules which prevent the occurence
of such conflicts. The latter solution, by imposing delay insensitivity
both on circuits specification and decomposition in a consistent way,
has the advantage of providing a purely logical characterization of the 
causal constraints. DI circuits then build a class which may be seen
as intermediate between synchronous circuits and general asynchronous
circuits. They share with the former
the possibility to be completely characterized by formal expressions
and rules. But they rely on the same causal structure as the latter. 
Synchronous circuits rely on time simultaneity classes,
and thus on a causal structure which is typical of Newtonian space-time.
Asynchronous circuits, on another hand, rely on a consistent treatment of 
propagation delays and time ordering, hence on a causal structure which
characterizes relativistic space-time. 

Delay-insensitivity provides an interesting transition between local
properties, like those defining sequential processors, and global 
ones, like those exhibited by distributed systems.
But DI circuits hardly exhaust the computation potentialities 
brought by the introduction
of asynchronous circuits. The critical consequences of
delay sensitivity rather suggest to consider a further alternative when
attempting to classify the different types of computations, i.e. those 
performed by
synchronous, by DI asynchronous and 
by DS (delay-sensitive) asynchronous circuits.
Similarly, 
in the same way as asynchronous computing machines may not
always allow simulation by synchronous computing machines,
one may infer that physical processes and physical laws, which
intrinsically obey relativistic causality, 
may be simulated by synchronous
machines in particular cases only. This hints at another advantage
of computations based on asynchronous circuits, i.e. the ability
to simulate in a universal way real physical processes.



\end{document}